\documentclass[
  journal=pasa,
  manuscript=research-paper, %% or "review"
  year=202X,
  volume=YY
]{cup-journal}

\usepackage[nopatch]{microtype}
\usepackage{booktabs}
\usepackage{rotating}  %used to rotate things like pictures or things you \put down
\usepackage{theorem}   %for mathematical theorems and lemmas
\usepackage{bm}  %bold fond inside equations, eg. $\bm{\alpha}$
\usepackage{amsmath,amsfonts,amssymb}
\usepackage{booktabs}  %Makes nice tables in books (sorta like RevTeX can)
\usepackage{pdfsync}   %used to sync pdf output inorder to flip from tex file to pdf and back
\usepackage{pdfpages}
\usepackage{float}
\usepackage{threeparttable}
\usepackage{hyperref}
\usepackage{tcolorbox}
\usepackage{multirow}
\usepackage{silence}
\newcommand\ion[2]{\text{#1\,\textsc{\lowercase{#2}}}}	% ionisation states
\title{Abundance Analysis of Chemically Depleted Post-AGB/Post-RGB Binaries with Faint Discs}

\author{Maksym Mohorian}
\affiliation{School of Mathematical and Physical Sciences, Macquarie University, Balaclava Road, Sydney, NSW 2109, Australia} 
\alsoaffiliation{Astrophysics and Space Technologies Research Centre, Macquarie University, Balaclava Road, Sydney, NSW 2109, Australia}
\email[Maksym Mohorian]{maksym.mohorian@students.mq.edu.au}

\author{Devika Kamath}
\affiliation{School of Mathematical and Physical Sciences, Macquarie University, Balaclava Road, Sydney, NSW 2109, Australia} 
\alsoaffiliation{Astrophysics and Space Technologies Research Centre, Macquarie University, Balaclava Road, Sydney, NSW 2109, Australia}
\alsoaffiliation{INAF, Osservatorio Astronomico di Roma, Via Frascati 33, I-00077 Monte Porzio Catone, Italy}
\author{Meghna Menon}
\affiliation{School of Mathematical and Physical Sciences, Macquarie University, Balaclava Road, Sydney, NSW 2109, Australia}
\alsoaffiliation{Astrophysics and Space Technologies Research Centre, Macquarie University, Balaclava Road, Sydney, NSW 2109, Australia}

\author{Hans Van Winckel}
\affiliation{Institute of Astronomy, KU Leuven, Celestijnenlaan 200D, 3001 Leuven, Belgium}

\author{Mingjie Jian}
\affiliation{Department of Astronomy, Stockholm University, AlbaNova University Center, Roslagstullsbacken 21, 114 21 Stockholm, Sweden}

\author{Anish M. Amarsi}
\affiliation{Theoretical Astrophysics, Department of Physics and Astronomy, Uppsala University, Box 516, SE-751 20 Uppsala, Sweden}

\author{Kateryna Andrych}
\affiliation{School of Mathematical and Physical Sciences, Macquarie University, Balaclava Road, Sydney, NSW 2109, Australia}
\alsoaffiliation{Astrophysics and Space Technologies Research Centre, Macquarie University, Balaclava Road, Sydney, NSW 2109, Australia}

\doi{ZZZZZ}

\received {dd Mmm 202Y}
\revised  {dd Mmm 202Y}
\accepted {dd Mmm 202Y}
\published{dd Mmm 202X}

 %% First letter not capped
\begin{document}

\begin{abstract}
Post-AGB and post-RGB binaries with stable circumbinary discs provide key insights into late stellar and disc evolution, revealing how binary interactions shape disc structure and stellar surface composition. A defining trait of such systems is the observed underabundance of refractory elements in the stellar photosphere relative to volatile elements -- photospheric chemical depletion -- resulting from the star accreting volatile-rich circumstellar gas. In this study, we investigated the link between photospheric depletion and disc evolution by focusing on post-AGB/post-RGB binaries with low infrared excess (hereafter ``faint disc'' targets). We analysed high-resolution optical spectra from HERMES/Mercator and UVES/VLT for 6 Galactic and 2 LMC targets. Using \texttt{E-iSpec}, we homogeneously derived atmospheric parameters and chemical abundances of 29 elements from carbon to europium, and included NLTE corrections for 15 elements from carbon to barium that we calculated using pySME and pre-computed grids of departure coefficients. All targets exhibit `saturated' depletion patterns, which we characterised using two-piece linear fits defined by three parameters: initial metallicity ([M/H]$_0$), turn-off temperature ($T_{\rm turn-off}$), and depletion scale ($\nabla_{\rm 100\,K}$). Among several findings, we highlight the bimodal distribution of $T_{\rm turn-off}$ in faint disc targets, which allows classification into two subgroups analogous to full discs with continuous, optically thick dust ($T_{\rm turn-off}\,>\,1\,100$\,K), and transition discs with inner clearing ($T_{\rm turn-off}\,<\,1\,100$\,K). Our results imply that faint disc targets likely represent the final stages of disc dissipation, highlighting the diversity of depletion profiles, the complexity of disc-binary interactions, and the need to understand the rarity and evolution of faint disc systems.
\end{abstract}

\section{Introduction}\label{sec:int}
Post-asymptotic giant branch (post-AGB) and post-red giant branch (post-RGB) binaries with circumbinary discs (CBDs) offer a unique opportunity to study disc-binary interactions and their impact on stellar surface chemistry. Toward the end of the AGB phase, low- and intermediate-mass stars ($0.8\,-\,8\,M_\odot$) lose most of their envelope mass through stellar winds \citep{vanwinckel2003Review, kobayashi2020OriginOfElements, kamath2023models}. In binary systems, the AGB/RGB phase can be prematurely terminated by mass loss, enhanced by poorly understood binary interactions \citep{paczynski1971RLOF, vanwinckel2003Review}. This process leads to the formation of a post-AGB or post-RGB primary star and the surrounding CBD \citep{vanwinckel2003Review, kamath2016PostRGBDiscovery, vanwinckel2018Binaries}.

The presence of CBDs around post-AGB and post-RGB binaries has been observationally confirmed through the detection of dust excess, which manifests as a broad near-infrared (near-IR) excess in their spectral energy distributions (SEDs) \citep{deruyter2005discs, kamath2014SMC, kamath2015LMC, kamath2016PostRGBDiscovery, gezer2015WISERVTau, kluska2022GalacticBinaries}. Detailed imaging studies have revealed complex dust structures within these discs, including cavities, rings, and arc-like features, through interferometric \citep{kluska2019DiscSurvey, corporaal2023DiscParameters} and polarimetric techniques \citep{ertel2019Imaging, andrych2023Polarimetry, andrych2024IRAS08}. Additionally, the outer regions of CBDs exhibit signatures of dust crystallisation \citep{gielen2011silicates, hillen2015ACHerMinerals}, grain growth \citep{scicluna2020GrainGrowth}, and Keplerian rotation \citep{bujarrabal2015KeplerianRotation, bujarrabal2018IRAS08, gallardocava2021PostAGBOutflows, gallardocava202389HerNebula, alcolea2023RedRectangle}.

\citet{kluska2022GalacticBinaries} classified Galactic post-AGB/post-RGB binaries into five categories, depending on their near-IR $H-K$ and mid-IR $W_1-W_3$ colours. Based on the IR colour-colour plot, targets with moderate mid-IR excess ($2.3^m<W_1-W_3<4.5^m$) potentially host full discs, where dust distribution starts from the sublimation radius \citep{hillen2016IRAS08, hillen2017DiscInterferometry, kluska2018IRAS08}. In contrast, targets with high mid-IR excess ($W_1-W_3>4.5^m$) and/or low near-IR excess ($H-K<0.4^m$) potentially host transition discs, which display inner gaps or cavities in the dust distribution \citep{corporaal2023DiscParameters}. Finally, targets with low IR excess ($H-K<0.4^m$, $W_1-W_3<2.3^m$) do not clearly fit into the full or transition disc categories and remain less understood. Despite differences in IR excess, all categories show evidence of disc–binary interactions, although the mechanisms driving these interactions remain poorly understood \citep{heath2020DiscBinaryInteraction, penzlin2022DiscBinaryInteraction}. Observational signatures of these interactions include jet formation \citep[results from interaction between the disc and the companion star, which is typically a main-sequence star][]{oomen2018OrbitalParameters, verhamme2024DiscWindModelling, deprins2024Jets} and photospheric chemical depletion \citep[arises from interaction between the disc and the post-AGB/post-RGB primary][]{oomen2019depletion, oomen2020MESAdepletion}.

In this study, we focus on photospheric chemical depletion, which occurs when gas in the CBD becomes fractionated from dust and is subsequently re-accreted onto the post-AGB/post-RGB star, altering its observed surface composition \citep{vanwinckel1995ExtremelyDepletedPostAGB, vanwinckel1997depletion, maas2005DiscPAGBs, giridhar2005rvtau, gezer2015WISERVTau, oomen2019depletion}. The efficiency of gas-dust fractionation depends on the condensation properties of the gas mixture, commonly described by 50\% condensation temperature $T_{\rm cond,\,50\%}$\footnote{At 50\% condensation temperature for a selected element, half of the element by abundance is in the gas phase, while another half is in the dust phase. These condensation temperatures are commonly calculated for solar mixture gas in chemical equilibrium at pressure $P\,=\,10^{-4}$\,bar, so these temperatures should not be considered absolute. Nevertheless, all targets analysed in this study are oxygen-rich ($0.07<$C/O$<0.81$; see Section~\ref{ssec:anaspc}), with C/O ratios close to the solar value \citep[C/O$_\odot\sim0.6$;][]{asplund2021solar}, implying that the major condensates should be the same \citep[e.g., corundum, olivines, pyroxenes;][]{agundez2020AGBDepletionModelling}.} \citep[$T_{\rm cond}$;][]{lodders2003CondensationTemperatures, wood2019CondensationTemperatures, agundez2020AGBDepletionModelling}. The fractionation process in the CBD separates gas containing elements with condensation temperatures $T_{\rm cond}\,\lesssim\,1\,250$\,K (volatile elements, including Na, S, Cu, and Zn) and dust containing elements with condensation temperatures $T_{\rm cond}\,\gtrsim\,1\,250$\,K (refractory elements, including Al, Si, Ti, Fe, and neutron capture process elements). Consequently, re-accreted volatile-rich matter dominates the original stellar surface composition, completely masking the nucleosynthetic products of AGB/RGB evolution \citep{deruyter2005discs, deruyter2006discs, oomen2018OrbitalParameters}. The resulting photospheric underabundance (depletion) of refractory elements in the post-AGB/post-RGB binaries is particularly prominent in the relative [X/H] abundance scale and is commonly plotted against $T_{\rm cond}$ \citep[depletion profile;][]{deruyter2005discs, maas2005DiscPAGBs, oomen2019depletion}.

Photospheric depletion in post-AGB/post-RGB binaries is commonly explored using four key parameters \citep{vanwinckel2018Binaries, kluska2022GalacticBinaries}:
\begin{itemize}
    \item The turn-off temperature $T_{\rm turn-off}$, which marks the separation between weakly depleted and significantly depleted elements. In previous studies, this parameter was estimated visually \citep{oomen2019depletion, kluska2022GalacticBinaries}.
    \item The initial metallicity [M/H]$_0$ ([S/H], [Zn/H], or their average, since Fe is depleted in post-AGB/post-RGB binaries), which traces the baseline composition of metals in the stellar surface. We note that volatile elements (including O, S, and Zn) may be mildly depleted \citep{mohorian2024EiSpec, mohorian2025TransitionDiscs}.
    \item The volatile-to-refractory abundance ratio ([Zn/Ti], [Zn/Fe], or [S/Ti]), which provides a scale for the efficiency of depletion \citep{waelkens1991depletion, oomen2019depletion, oomen2020MESAdepletion}. Based on [Zn/Ti] abundance ratios, post-AGB/post-RGB binaries can be categorised into four distinct groups: i) mildly depleted ([Zn/Ti]$\,\lesssim\,0.5$\,dex), ii) moderately depleted (0.5\,dex$\,\lesssim\,$[Zn/Ti]$\,\lesssim\,1.5$\,dex), iii) strongly depleted (1.5\,dex$\,\lesssim\,$[Zn/Ti]$\,\lesssim\,2.5$\,dex), and iv) extremely depleted ([Zn/Ti]$\,\gtrsim\,2.5$\,dex). We note that [Zn/Ti] abundance ratio depends on $T_{\rm turn-off}$ and [M/H]$_0$.
    \item The abundance pattern of the most refractory elements ($T_{\rm cond}\,\gtrsim\,$1\,400\,K): i) linear decline of [X/H] with increasing $T_{\rm cond}$ (`saturation') or ii) near-constant [X/H] with increasing $T_{\rm cond}$ (`plateau'). The saturated and plateau profiles represent systems where the re-accreted matter is fully or partially diluted in the stellar photosphere, respectively \citep{oomen2019depletion}.
    %\item The pattern of high-temperature end of the [X/H] depletion profile: i) linear decline of [X/H] with increasing $T_{\rm cond}$ (`saturation') or ii) constant [X/H] with increasing $T_{\rm cond}$ (`plateau'). This difference between saturated and plateau profiles is hypothesised to stem from the timescale of dilution of re-accreted matter in the stellar photosphere \citep{oomen2019depletion}.
\end{itemize}

CNO elements (C, nitrogen N, and oxygen O) are commonly omitted from the depletion analyses of post-AGB/post-RGB binaries, since the surface abundances of CNO elements are significantly altered by nucleosynthetic and mixing processes during the AGB/RGB evolution \citep{oomen2019depletion, oomen2020MESAdepletion, mohorian2024EiSpec, menon2024EvolvedBinaries}. For this reason, we similarly excluded CNO elements from our depletion analysis of post-AGB and post-RGB binaries.

The depletion efficiency of non-CNO elements is highly diverse across the post-AGB/post-RGB sample with different subtypes of CBDs: full disc targets generally show milder depletion ([Zn/Ti]\,$\lesssim$ 1 dex), transition disc targets commonly display the strongest depletion ([Zn/Ti]\,$\gtrsim$ 2 dex), and faint disc targets display a wide range of depletion efficiencies. However, we note that a small subset of full disc targets displays extremely low metallicity ([Fe/H] $\lesssim$ --3\,dex). This subset includes Red Rectangle, AG Ant, PS Gem, and HP Lyr \citep{kluska2022GalacticBinaries}, and the physical mechanisms behind the extreme depletion of Fe and other refractory elements in these full disc targets remain uncertain.

Furthermore, photospheric depletion is not limited to post-AGB and post-RGB binaries; a similar phenomenon was observed in young planet-hosting stars \citep[$\lambda$ B\"{o}o phenomenon;][]{venn1990lambdaBooStars, andrievsky2002lambdaBooAbundances, jura2015lambdaBoo, murphy2020lambdaBoo}. The depletion profiles in stars showing $\lambda$ B\"{o}o phenomenon suggest a possible connection between depletion mechanisms and the processes of planet formation \citep{kama2015DiscDepletionLinkinYSOs, jermyn2018Depletion}. Similarly, the depletion profiles in post-AGB/post-RGB binaries with transition discs were proposed to stem from second-generation planet formation in their CBDs \citep{kluska2022GalacticBinaries}.

To investigate how the diversity of photospheric depletion profiles is affected by the CBDs around post-AGB/post-RGB binaries, the nature of different disc types needs to be understood. In our previous studies, we homogeneously explored the depletion profiles of subsets of targets hosting full and transition discs \citep{mohorian2024EiSpec, mohorian2025TransitionDiscs}. In this study, we focus on a subset of post-AGB and post-RGB binaries with low IR excess, corresponding to category 4 in the classification scheme of \citet{kluska2022GalacticBinaries}. We refer to these as faint disc targets. Unlike full and transition disc targets, faint disc systems lack strong near- or mid-infrared excess, yet still show chemical evidence of disc-binary interaction. By analysing the chemical abundance profiles of faint disc targets and comparing them with previously studied full and transition disc systems, we aim to place faint disc targets within an evolutionary context. We propose three possible explanations for why these discs appear faint: i) a reduced pressure scale height, suggesting gas-poor conditions due to the settling of large grains into the midplane, and ii) a reduced dust content, as IR excess alone does not directly trace the total dust mass, or iii) a combination of dust settling in the midplane and dust reduction in the disc \citep{vanwinckel2003Review, vanwinckel2009, tosi2022DustContent, dellagli2023silicates}.

The structure of the paper is as follows. In Section~\ref{sec:sdo}, we describe our target sample, data, and observation details. In Section~\ref{sec:ana}, we outline the methodology and results of our chemical analysis of faint disc targets. In Section~\ref{sec:dsc}, we discuss photospheric depletion in faint disc targets and its connection to CBD evolution. Finally, in Section~\ref{sec:con}, we summarise the key findings of this study.

\section{Sample, data, and observations}\label{sec:sdo}
In this study, we focus on a subset of post-AGB/post-RGB binary stars with faint discs. In Section~\ref{ssec:sdosam}, we present the target sample and explain the selection criteria. In Section~\ref{ssec:sdopht}, we present the photometric data used to derive luminosities of our target sample (using SED fitting and PLC relation; see Section~\ref{ssec:analum}). In Section~\ref{ssec:sdospc}, we briefly discuss the spectroscopic data used to derive precise atmospheric parameters and elemental abundances of the faint disc targets (using \texttt{E-iSpec} and \texttt{pySME}; see Section~\ref{ssec:anaspc}).

\subsection{Sample selection}\label{ssec:sdosam}
The initial target sample consisted of post-AGB and post-RGB binary stars in the Galaxy \citep{kluska2022GalacticBinaries} and the Magellanic Clouds \citep[from][]{kamath2014SMC, kamath2015LMC}. The binarity of the targets in the initial sample was either established through orbital solutions derived from long-term spectroscopic monitoring or indirectly inferred from chemical depletion patterns, which are characteristic signatures of binary interaction \citep{vanwinckel2003Review, deruyter2006discs}. From this initial sample of 85 Galactic targets, 28 SMC targets, and 91 LMC targets, we selected faint disc targets based on their IR magnitudes from the 2MASS Long Exposure (6X) Full Survey \citep{cutri20122MASS6X} and the AllWISE catalogue \citep{cutri2014AllWISE}, using the selection criteria $H-K < 0.4^m$ and $W_1 - W_3 < 2.3^m$ adopted from \citet{kluska2022GalacticBinaries}. 

We further refined our target selection by only retaining objects with spectral types A to K for which we had high-resolution optical spectra (see Section~\ref{ssec:anaspc}). These spectral types allow for the optimal identification of \ion{Fe}{i} and \ion{Fe}{ii} spectral features. As a result, BD+15\,2862 was removed from our sample \citep[spectral type B9, $T_{\rm eff}\sim12\,000$\,K;][]{giridhar2005HotFaint}. Additionally, we excluded an LMC object, MACHO 47.2496.8 (OGLE LMC-T2CEP-015), due to the observed enhancement of slow neutron capture process (\textit{s}-process) elements in this target \citep{reyniers2007MeghnaMACHO, kamath2014SMC, menon2024EvolvedBinaries}. \citet{menon2024EvolvedBinaries} suggested that the depletion process (characteristic of our sample) is inhibited or absent in MACHO 47.2496.8, indicating that this target follows a different evolutionary pathway and is therefore not suitable for our study. As a result, our final target sample consists of 8 post-AGB/post-RGB binaries -- 6 in the Galaxy and 2 in the LMC. For SS\,Gem, V382\,Aur, and BD+39\,4926, binary nature is confirmed by the measured orbital parameters (see Table~\ref{tab:litpar}). For 5 other faint disc targets, binarity is inferred based on the observed chemical depletion patterns -- a well-established signature of binarity in post-AGB/post-RGB stars (see Section~\ref{sec:int}). In addition, the [Zn/Ti] abundance ratios within the sample \citep[commonly used as a tracer of depletion efficiency;][]{vanwinckel1992depletion, vanwinckel2003Review, deruyter2006discs, gielen2009Depletion, gorlova2012BD+46442, oomen2019depletion, kluska2022GalacticBinaries} vary from 0.2\,dex for R\,Sct to $\gtrsim\,3$\,dex for CC\,Lyr.

\begin{figure*}[!ht]
    \centering
    \includegraphics[width=0.75\linewidth]{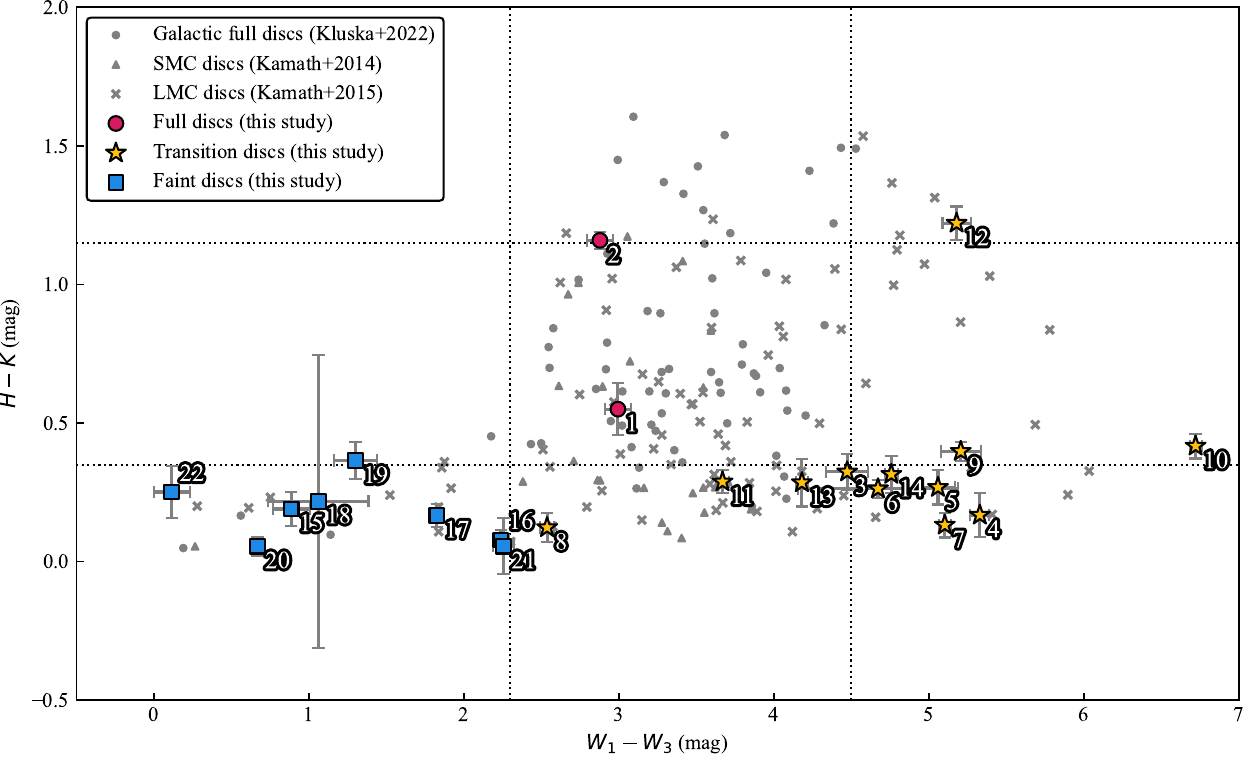}
    \caption[Updated IR colour-colour plot of post-AGB/post-RGB binary stars in the Galaxy and in the Magellanic Clouds]{Updated IR colour-colour plot of post-AGB/post-RGB binary stars in the Galaxy and in the Magellanic Clouds. NIR magnitudes ($H$ and $K$) are adopted from 2MASS 6X, while MIR magnitudes ($W_1$ and $W_3$) are adopted from AllWISE (for more details, see Section~\ref{ssec:sdopht}). Grey markers represent the extended sample of post-AGB/post-RGB binaries in the Galaxy \citep[circles;][]{kluska2022GalacticBinaries}, in the Small Magellanic Cloud \citep[triangles;][]{kamath2014SMC}, and in the Large Magellanic Cloud \citep[crosses;][]{kamath2015LMC}. Coloured markers represent the post-AGB/post-RGB binaries homogeneously studied with \texttt{E-iSpec+pySME} (for more details, see Section~\ref{ssec:dscall}): red circles mark the full disc targets \citep{mohorian2024EiSpec}, yellow stars mark the transition disc targets \citep{mohorian2025TransitionDiscs}, and blue squares mark the faint disc targets (this study). Adopted contours represent the demarcation between different categories of disc targets \citep[see Section~\ref{sec:int};][]{kluska2022GalacticBinaries}.}\label{fig:colplt}
\end{figure*}

We note that most of the faint disc targets (except for BD+39\,4926) are classified as RV\,Tauri pulsating variables. For all such targets, we derived luminosities using the period-luminosity-colour (PLC) relation (see Section~\ref{ssec:analum}). In Table~\ref{tab:sample}, we present the names, coordinates, and IR colours of our final sample of faint disc targets. In Figure~\ref{fig:colplt}, we show the IR colours of 8 faint disc targets from this study (marked by blue squares; for more details, see Section~\ref{ssec:dscall}). In Table~\ref{tab:litpar}, we provide relevant literature data for our faint disc sample, including orbital and pulsational parameters, luminosities, and depletion parameters (for more details, see \ref{app:tar}).

%\begin{figure*}[ph!]
%    \centering
%    \includegraphics[width=.845\linewidth]{figures/PDFs/FigColorPlot1.pdf}
%    \includegraphics[width=.845\linewidth]{figures/PDFs/FigColorPlot2.pdf}
%    \caption[Updated IR colour-colour plot of post-AGB/post-RGB binary stars in the Galaxy and in the Magellanic Clouds]{Updated IR colour-colour plot of post-AGB/post-RGB binary stars in the Galaxy and in the Magellanic Clouds. NIR magnitudes ($H$ and $K$) are adopted from 2MASS 6X, while MIR magnitudes ($W_1$ and $W_3$) are adopted from AllWISE (for more details, see Section~\ref{ssec:sdopht}). Grey markers represent the extended sample of {\color{red} post-AGB/post-RGB binaries} in the Galaxy \citep{kluska2022GalacticBinaries}, in the Small Magellanic Cloud \citep{kamath2014SMC}, and in the Large Magellanic Cloud \citep{kamath2015LMC}. Markers coloured by turn-off temperature $T_{\rm turn-off}$ (upper panel) and depletion scale $\nabla_{\rm 100\,K}$ (lower panel) represent the {\color{red} post-AGB/post-RGB binaries homogeneously studied with \texttt{E-iSpec+pySME}} (for more details, see Section~\ref{ssec:dscall}): circles mark the full disc targets \citep{mohorian2024EiSpec}, stars mark the transition disc targets \citep{mohorian2025TransitionDiscs}, and squares mark the faint disc targets (this study). Adopted contours represent the demarcation between different categories of disc targets \citep[see Section~\ref{sec:int};][]{kluska2022GalacticBinaries}.}\label{fig:colplt}
%\end{figure*}

\subsection{Photometric data}\label{ssec:sdopht}
To construct SEDs, we compiled photometric magnitudes spanning the optical to far-IR wavelength range (see Table~\ref{tab:phomag}). This includes $UBVRI$ photometry in the Johnson-Cousins system \citep{johnson1953Filters, cousins1976Filters}; $B_T$ and $V_T$ bands from the Tycho-2 catalogue \citep{hog2000Tycho2}; $I$' band from the SDSS \citep{york2000SDSSphotometry}; $J$, $H$, and $K$ magnitudes from 2MASS \citep{skrutskie20062MASS}; mid-IR $W_1$ to $W_4$ bands from WISE \citep{wright2010WISE}, and far-IR fluxes from AKARI, IRAS, PACS, and SPIRE \citep{ishihara2010AKARI, neugebauer1984IRAS, poglitsch2010PACS, griffin2010SPIRE}. For BD+28\,772, ultraviolet magnitudes from 156.5 nm to 274 nm were also adopted from the TD1 catalogue \citep{thompson1978TD1catalogue}.

\begin{table*}[!ht]%[ph!]
    \centering
    \footnotesize
    \caption[General details regarding faint disc sample (names, coordinates, and photometric selection criteria; see Section~\ref{ssec:sdosam})]{General details regarding faint disc sample (names, coordinates, and photometric selection criteria; see Section~\ref{ssec:sdosam}). $H$ and $K$ magnitudes originate from the 2MASS 6X catalogue, while the $W_1$ and $W_3$ magnitudes are sourced from the AllWISE catalogue (see Section~\ref{ssec:sdopht}).} \label{tab:sample}
    \begin{tabular}{|c|c|c|c|c|c|c|c|}
    \hline
        \multicolumn{3}{
        |c|}{\textbf{Names}} & \multicolumn{2}{c|}{\textbf{Coordinates}} & \multicolumn{2}{c|}{\textbf{Selection criteria}} \\ \hline
        \textbf{Adopted} & \textbf{IRAS/OGLE} & \textbf{2MASS} & \textbf{R.A.} & \textbf{Dec.} & \boldmath$H-K$ & \boldmath$W_1-W_3$ \\
        ~ & ~ & ~ & \textbf{(deg)} & \textbf{(deg)} & \textbf{(mag)} & \textbf{(mag)} \\\hline
        \multicolumn{7}{|c|}{\textit{Galactic targets}} \\\hline
        SS Gem & 06054+2237 & J06083510+2237020 & 092.146283 & +22.617229 & 0.190$\,\pm\,$0.060 & 0.886$\,\pm\,$0.119 \\
        V382 Aur & 06338+5333 & J06375242+5331020 & 099.468426 & +53.517227 & 0.078$\,\pm\,$0.035 & 2.239$\,\pm\,$0.050 \\
        CC Lyr & - & J18335741+3138241 & 278.489225 & +31.640045 & 0.167$\,\pm\,$0.041 & 1.826$\,\pm\,$0.041 \\
        R Sct & 18448-0545 & J18472894-0542185 & 281.870622 & --05.705157 & 0.218$\,\pm\,$0.528 & 1.061$\,\pm\,$0.324 \\
        AU Vul & 20160+2734 & J20180588+2744035 & 304.524513 & +27.734322 & 0.364$\,\pm\,$0.066 & 1.301$\,\pm\,$0.136 \\
        BD+39 4926 & - & J22461123+4006262 & 341.546798 & +40.107304 & 0.056$\,\pm\,$0.034 & 0.668$\,\pm\,$0.040 \\ \hline
        \multicolumn{7}{|c|}{\textit{LMC targets}} \\\hline
        J052204 & LMC-LPV-46487 & J05220425-6915206 & 80.517725 & --69.255730 & 0.056$\,\pm\,$0.100 & 2.257$\,\pm\,$0.068 \\
        J053254 & LMC-T2CEP-149 & J05325445-6935131 & 83.226904 & --69.586983 & 0.085$\,\pm\,$0.067 & 0.749$\,\pm\,$0.083$^a$ \\ \hline
        %\multicolumn{7}{|c|}{\textit{Galactic candidate}} \\\hline
        %BD+28 772 & 05140+2851 & J05171546+2854180 & 79.314440 & +28.905008 & 0.225$\,\pm\,$0.272 & 0.012$\,\pm\,$0.227 \\ \hline
    \end{tabular}\\
    \textbf{Notes:} $^a$[3.6]-[8.0] from SAGE catalogue \citep{woods2011SAGE} was adopted as mid-IR colour for J053254. By scaling to WISE wavelengths ($\lambda_{W_1}\,=\,3.368\,\mu$m, $\lambda_{W_3}\,=\,12.082\,\mu$m), we estimate $W_1-W_3\,\sim\,1.5^m$ for this target, which constitutes as low mid-IR excess (below the cut-off magnitude of $2.3^m$; see Section~\ref{ssec:sdosam}).
\end{table*}
\begin{table*}[!ht]%[ph!]
    \centering
    \footnotesize
    \caption{Parameters of faint disc sample, including orbital, pulsational, and depletion parameters, as well as luminosity estimates adopted from the literature (see Section~\ref{sec:sdo}).} \label{tab:litpar}
    \begin{tabular}{|c|cc|cc|cc|ccc|}
    \hline
        & \multicolumn{2}{c|}{\textbf{Orbital parameters}} & \multicolumn{2}{c|}{\textbf{Pulsational parameters}} & \multicolumn{2}{c|}{\textbf{Luminosity estimates}} & \multicolumn{3}{c|}{\textbf{Depletion parameters}} \\
        \textbf{Name} & \boldmath$P_{\rm orb}$ & \boldmath$e$ & \boldmath$P_{\rm puls}$ & \textbf{RVb} & \boldmath$L_{\rm SED}$ & \boldmath$L_{\rm IR}/L_\ast$ & \textbf{[Zn/Ti]} & \boldmath$T_{\rm turn-off}$ & \textbf{Profile} \\
        ~ & \textbf{(d)} & ~ & \textbf{(d)} & ~ & \boldmath$(L_\odot)$ & ~ & \textbf{(dex)} & \textbf{(K)} & \textbf{pattern} \\\hline
        SS Gem & 910$^o$ & -- & 44.525$^a$ & yes & 4540$\,\pm\,$1190$^m$ & 0.01$^m$ & 2.02$^g$ & 1100 & S \\
        V382 Aur & 597.4$\,\pm\,$0.2$^b$ & 0.30$\,\pm\,$0.02$^b$ & 29.06$^c$ & yes & 3540$\,\pm\,$810$^m$ & 0.02$^m$ & 0.87$^h$ & 800 & P \\
        CC Lyr & -- & -- & 23.634$^d$ & no & 650$\,\pm\,$210$^m$ & 0.03$^m$ & $\gtrsim$3$^i$ & 1000 & S \\
        R Sct & -- & -- & 70.8$^e$ & no & 2830$\,\pm\,$580$^m$ & 0.03$^m$ & 0.17$^j$ & -- & N \\
        AU Vul & -- & -- & 71.55$^a$ & no & 4220$\,\pm\,$1160$^m$ & 0.06$^m$ & -- & -- & -- \\
        BD+39 4926 & 871.7$\,\pm\,$0.4$^b$ & 0.024$\,\pm\,$0.006$^b$ & -- & no & 6120$\,\pm\,$1870$^m$ & $>$0.00$^m$ & 2.03$^k$ & 1000 & S \\ \hline
        J052204 & -- & -- & 30.172$^f$ & no & 4170$\,\pm\,$970$^n$ & 0.02$^o$ & -- & -- & -- \\
        J053254 & -- & -- & 42.744$^f$ & no & 3210$\,\pm\,$1210$^n$ & $>$0.00$^o$ & 0.6$^l$ & -- & -- \\ \hline
        %BD+28 772 & -- & -- & -- & -- & -- & -- & -- & -- & -- \\ \hline
    \end{tabular}\\
    %\raggedright
    \textbf{Notes:} The parameters are adopted from the following studies: $^a$\citet{pawlak2019ASAS}, $^b$\citet{oomen2018OrbitalParameters}, $^c$\citet{hrivnak2008V382Aur}, $^d$\citet{zong2020LAMOST}, $^e$\citet{kalaee2019RSct}, $^f$\citet{soszynski2008OGLE}, $^g$\citet{gonzalez1997CTOri}, $^h$\citet{hrivnak2008V382Aur}, $^i$\citet{maas2007t2cep}, $^j$\citet{giridhar2000RSct}, $^k$\citet{rao2012BD+394926}, $^l$\citet{gielen2009Depletion}, $^m$\citet{kluska2022GalacticBinaries}, $^n$\citet{manick2018PLC}, $^o$This study. For BD+39\,4926 and J053254, the IR-to-stellar luminosity ratios are in the range $0<L_{\rm IR}/L_\ast<0.01$. $T_{\rm turn-off}$ and profile patterns are adopted from \citet{oomen2019depletion}: `S' means `saturated', `P' means `plateau', `N' means `not depleted'.\\
\end{table*}
\begin{table*}[!ht]%[ph!]
    \centering
    \footnotesize
    \caption[Photometric data of faint disc sample (see Section~\ref{ssec:sdopht})]{Photometric data of faint disc sample (see Section~\ref{ssec:sdopht}). The table includes units and central wavelengths (in $\mu$m) for each filter. This table is published in its entirety in the electronic edition of the paper. A portion is shown here for guidance regarding its form and content.}\label{tab:phomag}
    \begin{tabular}{|c|c|c|c|c|c|}\hline
        \textbf{Filter} & \textbf{...} & \textbf{JOHNSON.B (I/280B/ascc)} & \textbf{...} & \textbf{2MASS.J (II/246/out)} & \textbf{...} \\
        \textbf{Unit} & \textbf{...} & \textbf{mag} & \textbf{...} & \textbf{mag} & \textbf{...} \\
        \boldmath$\lambda$ \textbf{(\boldmath$\mu$m)} & \textbf{...} & \boldmath$0.443$ & \textbf{...} & \boldmath$1.239$ & \textbf{...} \\ \hline
        SS Gem & ... & $10.298\,\pm\,0.052$ & ... & $6.743\,\pm\,0.032$ & ... \\
        V382 Aur & ... & $9.557\,\pm\,0.019$ & ... & $7.805\,\pm\,0.020$ & ... \\
        CC Lyr & ... & $13.000\,\pm\,0.259$ & ... & $10.832\,\pm\,0.019$ & ... \\
        R Sct & ... & -- & ... & $2.824\,\pm\,0.274$ & ... \\
        AU Vul & ... & $12.312\,\pm\,0.165$ & ... & $7.183\,\pm\,0.026$ & ... \\
        %BD+39 4926 & -- & -- & ... & -- & -- \\
        ... & ... & ... & ... & ... & ... \\ \hline
        %BD+28 772 & $(4.3\,\pm\,1.5)\,\cdot\,10^{-13}$ & $(1.1\,\pm\,0.3)\,\cdot\,10^{-12}$ & ... & -- & -- \\ \hline
    \end{tabular}
\end{table*}

\subsection{Spectroscopic data}\label{ssec:sdospc}
In this subsection, we present the high-resolution optical spectra used in this study, obtained with HERMES/Mercator \citep{raskin2011hermes, gorlova2011HERMESprogram} and UVES/VLT \citep{dekker2000UVES}. In Table~\ref{tab:obslog}, we present the observational log with measured radial velocities and periodicities in the spectra. To determine the radial velocities for all spectral visits of each faint disc target, we used E-iSpec (see Section~\ref{ssec:anaspc}). To analyse periodic variations in the derived radial velocity curves, we first applied the Lomb-Scargle periodogram to identify significant periodic signals \citep{lomb1976Periodogram, scargle1982Periodogram}, and then refined these periods using the Lafler-Kinman method \citep{lafler1965Periodogram}.

To derive precise atmospheric parameters and elemental abundances for our faint disc targets, we selected spectra in the following way (see Table~\ref{tab:obslog}):
\begin{itemize}
    \item For targets observed with HERMES, we selected optical visits with the highest S/N ratios for each target.
    \item For targets observed with UVES, we combined all available spectral visits.
\end{itemize}

\begin{figure}[!ht]
    \centering
    \includegraphics[width=.99\linewidth]{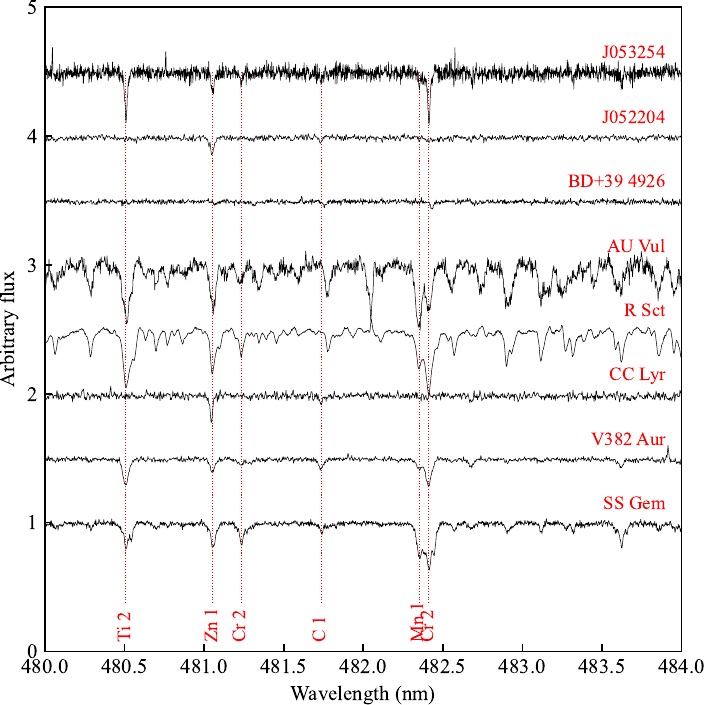}
    \caption[Spectra of all faint disc targets across the region featuring lines from the volatile S and Zn]{Spectra of all faint disc targets across the region featuring lines from the volatile S and Zn. Each spectrum is RV corrected, normalised, and offset in flux for clarity. Red dashed vertical lines mark the positions of spectral line peaks (for more details, see Section~\ref{ssec:anaspc}).}\label{fig:spcmon}
\end{figure}

In \ref{app:vis}, we provide details of all the spectral visits considered. In Figure~\ref{fig:spcmon}, we show the sample spectra of faint disc targets in the wavelength region near the S and Zn lines to illustrate the quality of our obtained data. We note that [S/H] and [Zn/H] abundances are commonly used to represent the initial metallicity [M/H]$_0$ of post-AGB/post-RGB binaries, rather than [Fe/H] abundance, due to the high efficiency of refractory depletion (see Section~\ref{sec:int}).

\begin{table}[ht]
    \centering
    \caption[Analysed spectral visits of faint disc targets]{Analysed spectral visits of faint disc targets. For more details on the criteria used for selecting spectral visits, see Section~\ref{ssec:sdospc}. For a complete observational summary of faint disc targets, see \ref{app:vis}.} \label{tab:obslog}
    \resizebox{\columnwidth}{!}{
    \begin{tabular}{|c|c|c|c|c|c|c|}
    \hline
        \textbf{Name} & \textbf{Facility} & \textbf{ObsID} & \textbf{MJD} & \textbf{RV (km/s)} & \textbf{S/N} & \boldmath$P_{\rm RV}$ \textbf{(d)} \\ \hline
        SS Gem & H/M & \begin{tabular}{c} 273787,\\273788,\\273789 \end{tabular} & \begin{tabular}{c} 55232.98736,\\55232.99957,\\55233.01175 \end{tabular} & --15.8 & 45 & 44.66 \\ \hline
        V382 Aur & H/M & \begin{tabular}{c} 904077,\\904090 \end{tabular} & \begin{tabular}{c} 58450.08702,\\58450.20395 \end{tabular} & --84.6 & 65 & 599.1 \\ \hline
        CC Lyr & H/M & \begin{tabular}{c} 356745,\\356746,\\356747 \end{tabular} & \begin{tabular}{c} 55723.94027,\\55723.96863,\\55723.99005 \end{tabular} & --35.8 & 55 & 23.69 \\ \hline
        R Sct & H/M & 972206 & 59073.92861 & 36.2 & 35 & 71.54 \\ \hline
        AU Vul & H/M & \begin{tabular}{c} 596805,\\596806 \end{tabular} & \begin{tabular}{c} 56935.86608,\\56935.88750 \end{tabular} & --2.1 & 35 & 75.84 \\ \hline
        BD+39 4926 & H/M & \begin{tabular}{c} 244722,\\244723 \end{tabular} & \begin{tabular}{c} 55067.06677,\\55067.08689 \end{tabular} & --15.8 & 80 & 879.0 \\ \hline
        J052204 & U/V & \begin{tabular}{c} 456+472,\\739+776 \end{tabular} & \begin{tabular}{c} 58451.31574,\\58470.18473 \end{tabular} & 231.3 & 45 & -- \\ \hline
        J053254 & U/V & 494 & 53409.20586 & 268.8 & 40 & -- \\ \hline
        %BD+28 772 & H/M & 427610 & 56210.06768 & --3.4 & 30 & 1596 \\ \hline
    \end{tabular}
    }\\
    \textbf{Notes:} H/M denotes HERMES/Mercator, and U/V denotes UVES/VLT. RV is the radial velocity of the spectrum used in the analyses (for merged spectral visits, the range of RVs is <0.5 km/s). S/N is the average signal-to-noise ratio of the spectrum. $P_{\rm RV}$ is the period derived from RV curves, initially estimated using a Lomb-Scargle periodogram and refined with the Lafler-Kinman method.
\end{table}

\section{Derivation of luminosities, atmospheric parameters, and elemental abundances of faint disc targets}\label{sec:ana}
In this section, we discuss the methodology and results of the photometric and spectral analysis of faint disc targets. In Section~\ref{ssec:analum}, we outline the derivation of luminosities using SED fitting and PLC relation, based on the collected photometric data. In Section~\ref{ssec:anaspc}, we present the derivation of precise atmospheric parameters and elemental abundances using high-resolution optical spectra. In Section~\ref{ssec:ananlt}, we describe the calculation of non-local thermodynamic equilibrium (statistical equilibrium, NLTE) corrections, which significantly impact the chemical analysis of post-AGB/post-RGB binaries. In Section~\ref{ssec:anares}, we present the depletion profiles of faint (this study), transition \citep{mohorian2025TransitionDiscs}, and full disc targets \citep{mohorian2024EiSpec}.

\subsection{Luminosity estimation using SED fitting and PLC relation}\label{ssec:analum}
In this study, we derived the luminosities of our target sample using the following methods: i) SED fitting to calculate the SED luminosities $L_{\rm SED}$ and ii) PLC relation to obtain the PLC luminosities $L_{\rm PLC}$. In this subsection, we detail the procedures used for each method.

\begin{figure*}[ph!]
    \centering
    \includegraphics[width=.415\linewidth]{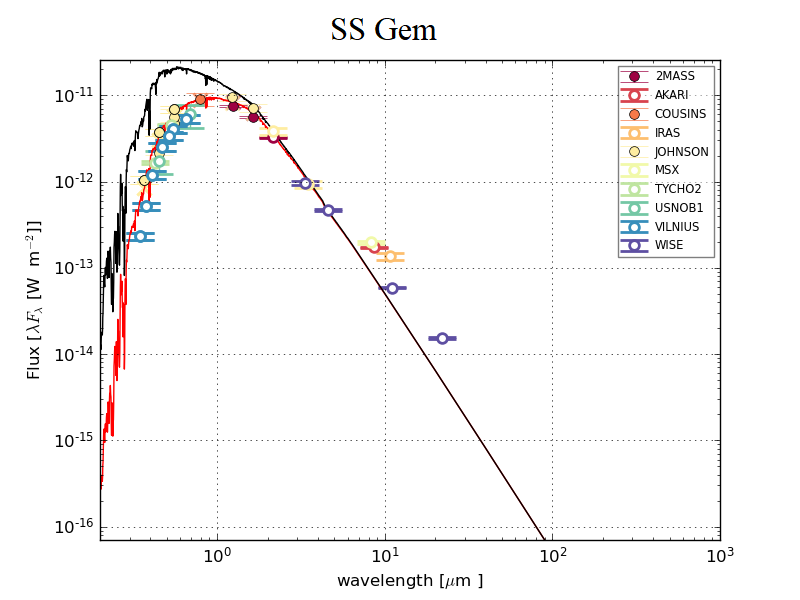}
    \includegraphics[width=.415\linewidth]{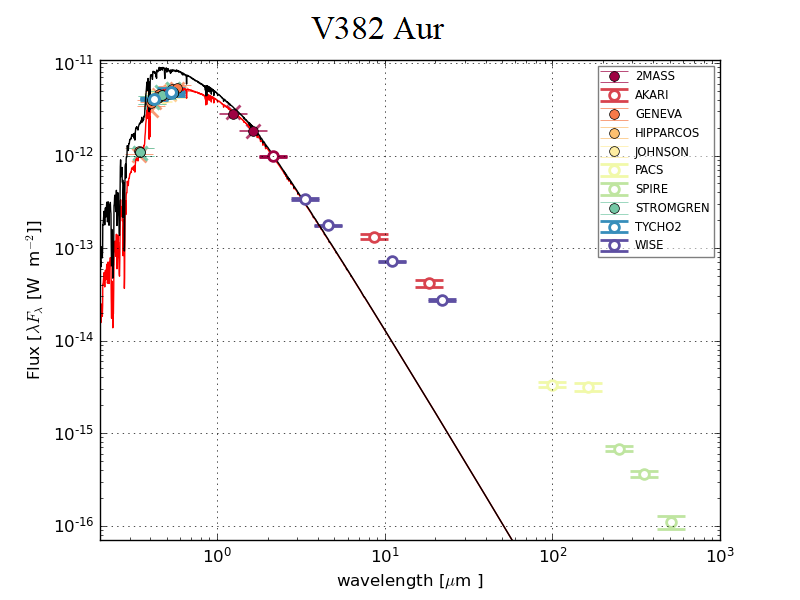}
    \includegraphics[width=.415\linewidth]{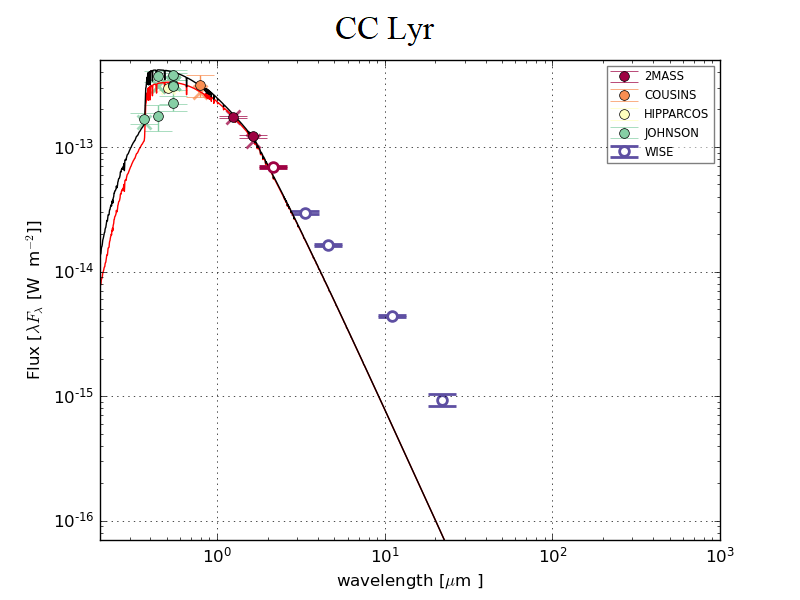}
    \includegraphics[width=.415\linewidth]{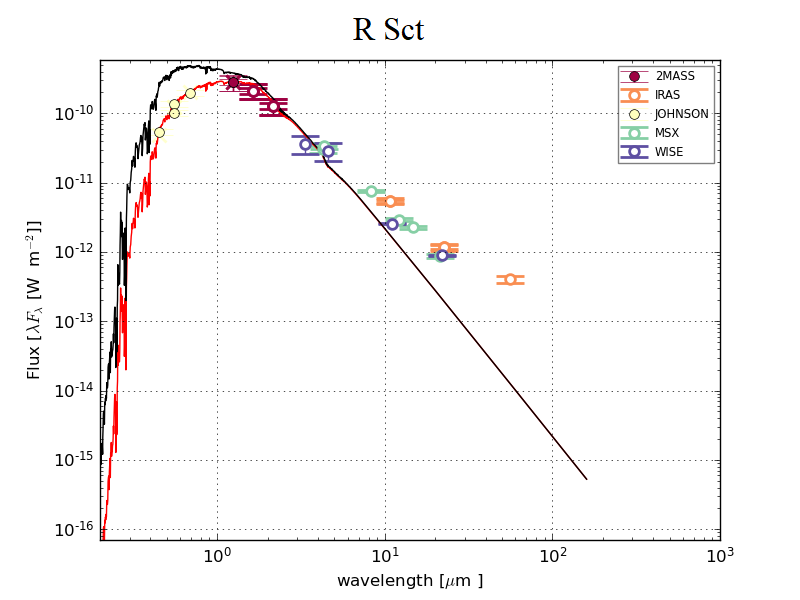}
    \includegraphics[width=.415\linewidth]{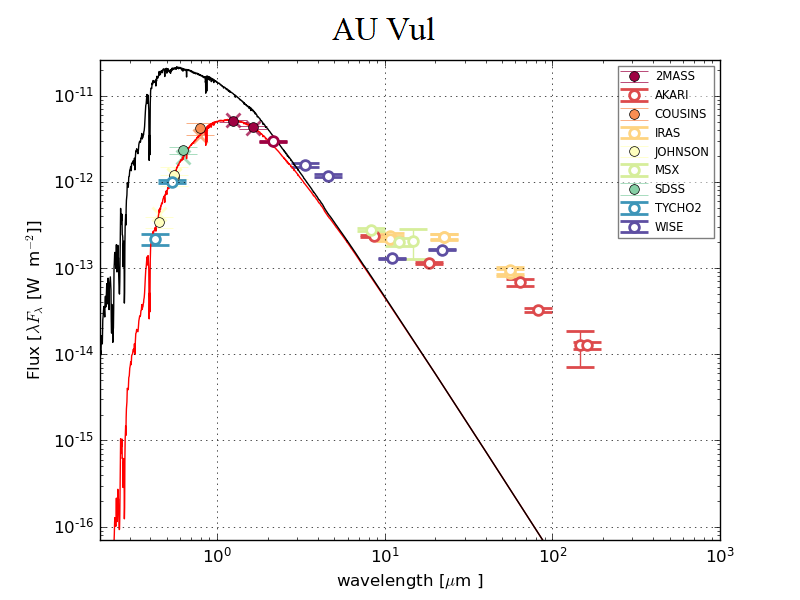}
    \includegraphics[width=.415\linewidth]{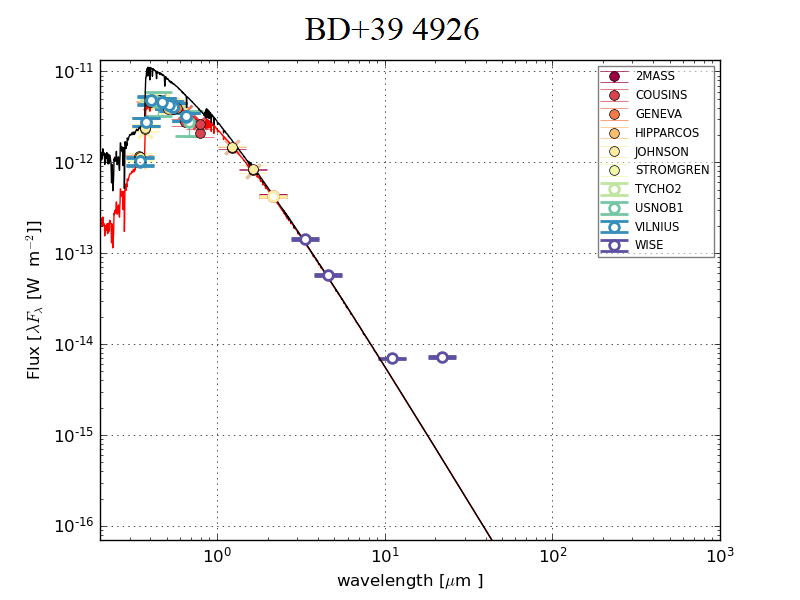}
    \includegraphics[width=.415\linewidth]{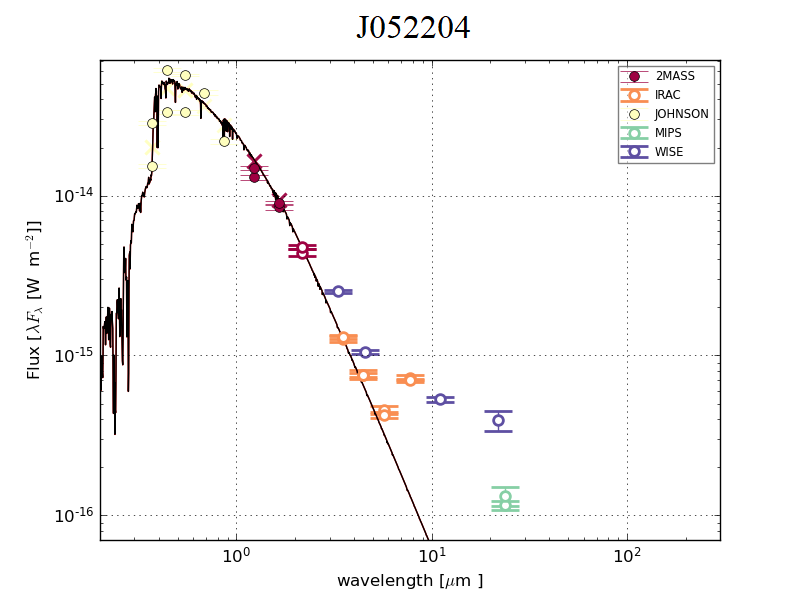}
    \includegraphics[width=.415\linewidth]{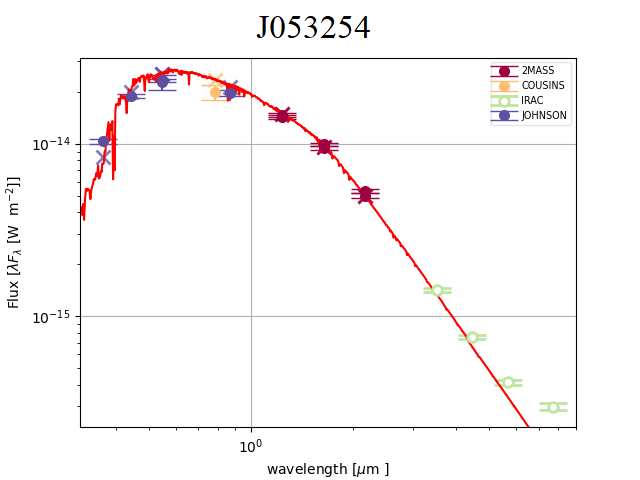}
    \caption[Spectral energy distribution plots of faint disc targets in the Galaxy and the LMC]{Spectral energy distribution plots of faint disc targets in the Galaxy and the LMC. The red solid curve corresponds to the reddened Kurucz model atmosphere, while the black solid curve represents Kurucz model atmosphere after de-reddening and scaling to the object. A legend within the plot clarifies the meaning of the symbols and colours used.}\label{fig:allSED}
    \vspace{-0.25cm}
\end{figure*}

To calculate $L_{\rm SED}$, we followed the procedure outlined in \citet{mohorian2024EiSpec}. In brief, we fitted photometric data with spectra synthesised from Kurucz ODFNEW model atmospheres \citep{castelli2003ATLAS9}, integrating the bolometric IR luminosity and correcting for total reddening \citep[interstellar and circumstellar; see][]{degroote2013SEDfitting}. $\chi^2$ minimisation was performed until the convergence of effective temperature $T_{\rm eff}$, surface gravity $\log g$, extinction (reddening) $E(B-V)$, and stellar angular size $\theta$. In our calculations, interstellar reddening follows the extinction law from \citet{cardelli1989SEDextinction} with $R_V\,=\,3.1^m$. We note that we used the Bailer-Jones geometric distances ($z_{\rm BJ}$) with \textit{Gaia} EDR3 parallaxes \citep{bailerjones2021distances}, assuming isotropic radiation emission. Additionally, we did not explicitly account for pulsational variability, which resulted in higher $\chi^2$ values for pulsating variables. The uncertainties in $L_{\rm SED}$ are mainly caused by uncertainties in distance and reddening. In Figure~\ref{fig:allSED}, we show the SEDs of faint disc targets.

We note that the IR excess of faint disc targets is not only weak in flux contribution ($0 < L_{\rm IR}/L_{\ast} < 0.1$), but also generally starts at $\sim10\,\mu$m (for V382 Aur and CC Lyr, at $\sim5\,\mu$m). This is significantly different from the IR excesses of full and transition disc systems, which start at $\sim1\,\mu$m and $\sim3\,\mu$m, respectively \citep{kluska2022GalacticBinaries}.

To calculate $L_{\rm PLC}$, we applied the calibrated relation described in \citet{menon2024EvolvedBinaries}, given by
\begin{equation}
    M_{bol} = m_{cal}\cdot\log P_0 + c_{cal} - \mu + BC + 2.55\cdot(V-I)_0,
\end{equation}
where $M_{bol}$ is the absolute bolometric magnitude; $m_{cal}=-3.59^m$ and $c_{cal}=18.79^m$ are the calibrated slope and intercept of PLC relation, respectively; $P_0$ is the observed fundamental period of pulsation; $\mu=18.49$ is the distance modulus to the LMC (used to calibrate the relation); $BC$ is the bolometric correction based on the $T_{\rm eff}$ \citep{flower1996BoloCorr, torres2010BoloCorrErrata}; and $(V-I)_0$ is the intrinsic (de-reddened) colour based on the reddening value $E(B-V)$ from the SED fit. The uncertainties in $L_{\rm PLC}$ are mainly caused by uncertainties in $E(B-V)$.

We note that $L_{\rm PLC}$ provides a more reliable luminosity estimate than $L_{\rm SED}$, since $L_{\rm SED}$ is more sensitive to model fitting variations caused by RV Tau pulsations \citep{menon2024EvolvedBinaries}. Therefore, for all pulsating variables in the faint disc sample, we use $L_{\rm PLC}$ as the adopted luminosity for our subsequent analysis. For the non-pulsating star, BD+39\,4926, we adopt $L_{\rm SED}$ instead. In Table~\ref{tab:respar}, we provide the derived values and uncertainties for both $L_{\rm SED}$ and $L_{\rm PLC}$.

%\begin{figure*}[!ht]
%    \centering
%    \includegraphics[width=.45\linewidth]{figures/PDFs/SEDs/J052204.24-691520.7SED.png}
%    \includegraphics[width=.45\linewidth]{figures/PDFs/SEDs/J053254.50-693513.2SED.png}
%    %\includegraphics[width=.44\linewidth]{figures/PDFs/SEDs/BD+28_772SED.png}
%    \caption[Spectral energy distribution plots of LMC faint disc stars]{Spectral energy distribution plots of LMC faint disc stars. The red solid curve corresponds to the reddened Kurucz model atmosphere, while the black solid curve represents Kurucz model atmosphere after de-reddening and scaling to the object. A legend within the plot clarifies the meaning of the symbols and colours used.}\label{fig:allSED2}
%    \vspace{-0.25cm}
%\end{figure*}
\begin{table*}[!ht]%[ph!]
    \centering
    \scriptsize
    \caption[Derived luminosities, atmospheric parameters, selected abundances and abundance ratios of faint disc targets (see Sections~\ref{ssec:analum} and \ref{ssec:anaspc})]{Derived luminosities, atmospheric parameters, selected abundances and abundance ratios of faint disc targets (see Sections~\ref{ssec:analum} and \ref{ssec:anaspc}). For a full list of [X/H] abundances, see \ref{app:abu}.}\label{tab:respar}
    \resizebox{\columnwidth}{!}{
    \begin{tabular}{|c|cc|cccc|cc|cc|}\hline
        ~ & \multicolumn{2}{c|}{\textbf{Derived luminosities}} & \multicolumn{4}{c|}{\textbf{Atmospheric parameters}} & \multicolumn{2}{|c|}{\textbf{LTE ratios}} & \multicolumn{2}{|c|}{\textbf{NLTE ratios}} \\
        \textbf{Target} & \boldmath$\log\dfrac{L_{\rm SED}}{L_\odot}$ & \boldmath$\log\dfrac{L_{\rm PLC}}{L_\odot}$ & \boldmath$T_{\rm eff}$ & \boldmath$\log g$ & \textbf{[Fe/H]} & \boldmath$\xi_{\rm t}$ & \textbf{[Zn/Ti]} & \textbf{[Zn/Fe]} & \textbf{[S/Ti]} & \textbf{C/O} \\
        &&& \textbf{(K)} & \textbf{(dex)} & \textbf{(dex)} & \textbf{(km/s)} & \textbf{(dex)} & \textbf{(dex)} & \textbf{(dex)} & ~ \\ \hline
        SS Gem & $3.657\,\pm\,0.116$ & $3.400_{-0.296}^{+0.099}$ & $6500\,\pm\,90$ & $1.96\,\pm\,0.11$ & $-1.02\,\pm\,0.12$ & $4.08\,\pm\,0.09$ & $0.98\,\pm\,0.21$ & $0.45\,\pm\,0.25$ & $1.25\,\pm\,0.09$ & $0.55\,\pm\,0.28$ \\
        V382 Aur & $3.549\,\pm\,0.101$ & $3.350_{-0.183}^{+0.113}$ & $6020\,\pm\,150$ & $0.38\,\pm\,0.27$ & $-1.82\,\pm\,0.08$ & $4.14\,\pm\,0.13$ & $0.81\,\pm\,0.34$ & $0.69\,\pm\,0.31$ & $1.64\,\pm\,0.19$ & $0.38\,\pm\,0.13$ \\
        CC Lyr & $2.811\,\pm\,0.143$ & $2.986_{-0.099}^{+0.211}$ & $6470\,\pm\,250$ & $1.64\,\pm\,0.50$ & $-3.80\,\pm\,0.03$ & $4.14\,\pm\,1.00$ & $-$ & $2.95\,\pm\,0.44$ & $-$ & $0.03\,\pm\,0.02$ \\
        R Sct & $3.452\,\pm\,0.090$ & $3.099_{-0.394}^{+0.113}$ & $5630\,\pm\,70$ & $0.93\,\pm\,0.11$ & $-0.60\,\pm\,0.11$ & $2.93\,\pm\,0.03$ & $0.54\,\pm\,0.17$ & $0.42\,\pm\,0.17$ & $0.31\,\pm\,0.13$ & $0.17\,\pm\,0.04$ \\
        AU Vul & $3.626\,\pm\,0.123$ & $3.816_{-0.141}^{+0.253}$ & $4740\,\pm\,40$ & $0.00\,\pm\,0.09$ & $-0.95\,\pm\,0.08$ & $4.32\,\pm\,0.04$ & $0.40\,\pm\,0.27$ & $0.08\,\pm\,0.36$ & $-$ & $0.07\,\pm\,0.00$ \\
        BD+39 4926 & $3.787\,\pm\,0.137$ & $-$ & $7470\,\pm\,40$ & $0.52\,\pm\,0.05$ & $-2.37\,\pm\,0.18$ & $2.21\,\pm\,0.25$ & $2.69\,\pm\,0.44$ & $2.17\,\pm\,0.29$ & $3.36\,\pm\,0.35$ & $0.14\,\pm\,0.10$ \\ \hline
        J052204 & $3.620\,\pm\,0.091$ & $3.307_{-0.000}^{+0.225}$ & $7020\,\pm\,100$ & $1.65\,\pm\,0.11$ & $-2.49\,\pm\,0.10$ & $2.36\,\pm\,0.04$ & $3.15\,\pm\,0.34$ & $1.98\,\pm\,0.33$ & $-$ & $0.54\,\pm\,0.21$ \\
        J053254 & $3.507\,\pm\,0.139$ & $3.843_{-0.479}^{+0.070}$ & $6040\,\pm\,90$ & $0.47\,\pm\,0.13$ & $-1.66\,\pm\,0.09$ & $2.36\,\pm\,0.03$ & $0.87\,\pm\,0.27$ & $0.45\,\pm\,0.23$ & $1.31\,\pm\,0.15$ & $0.81\,\pm\,0.21$ \\ \hline
        %BD+28 772 & 3.502\,\pm\,0.483 & - & 6880\,\pm\,100 & 0.96\,\pm\,0.16 & -0.42\,\pm\,0.10 & 2.14\,\pm\,0.03 & -0.10\,\pm\,0.25 & -0.25\,\pm\,0.25 & 0.26\,\pm\,0.19 & 0.39\,\pm\,0.11 \\ \hline
    \end{tabular}
    }\\
    \textbf{Note:} We adopt $L_{\rm PLC}$ for all targets except BD+39 4926 and BD+28 772, for which we adopt $L_{\rm SED}$.
    \vspace{-0.5cm}
\end{table*}

\subsection{Derivation of atmospheric parameters and elemental abundances using \texttt{E-iSpec}}\label{ssec:anaspc}
In this study, we derived atmospheric parameters and elemental abundances using \texttt{E-iSpec}, a semi-automated spectral analysis code \citep{mohorian2024EiSpec}, specifically tailored for evolved stars with complex atmospheres. \texttt{E-iSpec} is a modified version of the iSpec spectral analysis code \citep{blancocuaresma2014, blancocuaresma2019}. The analysis within \texttt{E-iSpec} is conducted using the LTE \texttt{Moog} transfer code \citep[operating with equivalent width method,][]{sneden2012Moog}, the VALD3 line list \citep{kupka2011vald}, solar abundances from \citet{asplund2021solar}, and 1D plane-parallel ATLAS9 model atmospheres \citep{castelli2003ATLAS9}. In \ref{app:lst}, we present the final line list selected for the precise derivation of atmospheric parameters and elemental abundances.

For atmospheric parameters, we adopt the original iSpec method for computing uncertainties. For elemental abundances, we compute the uncertainties using a quadrature sum of random and systematic components \citep[for more details, see][]{mohorian2025TransitionDiscs}. We note that the random component is set to 0.1 dex for elements with abundance derived from a single spectral line. Additionally, we assume that variations in metallicity affect the [X/Fe] abundance ratios, but not the [X/H] abundances. 

\begin{figure*}
    \centering
    \includegraphics[width=.405\linewidth]{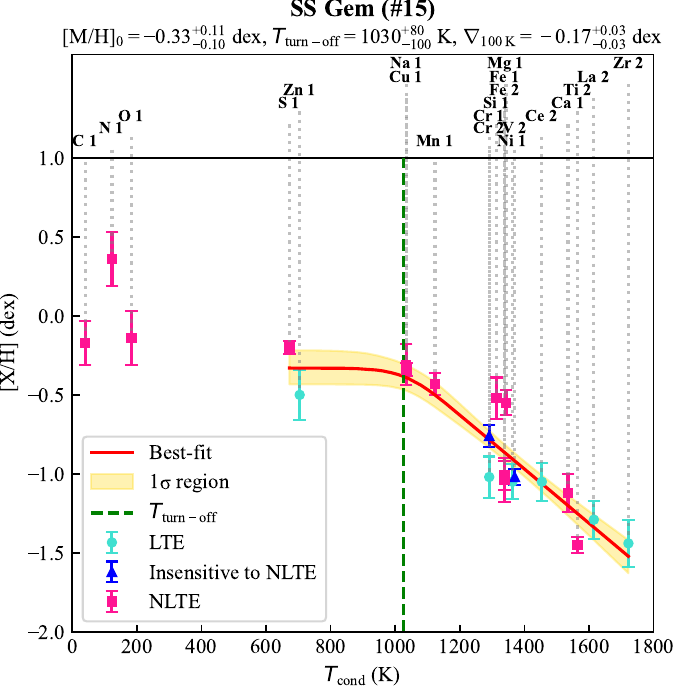}
    \includegraphics[width=.405\linewidth]{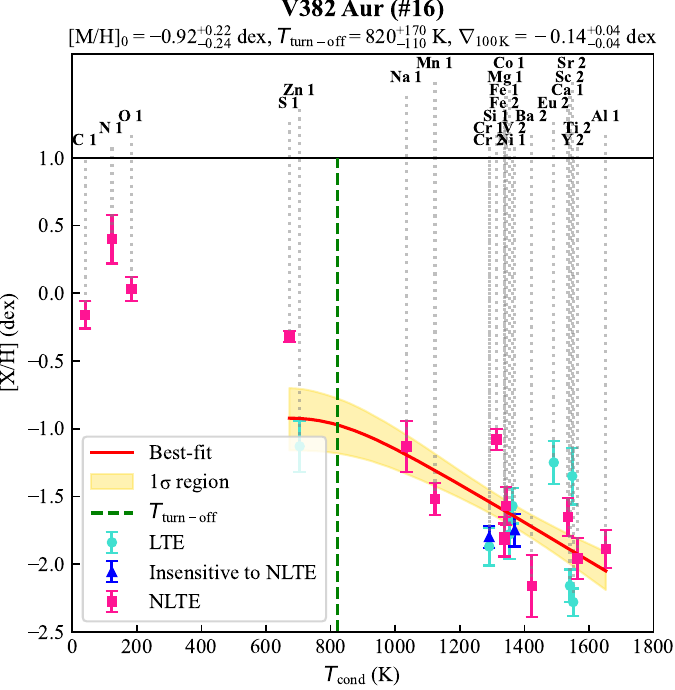}
    \includegraphics[width=.405\linewidth]{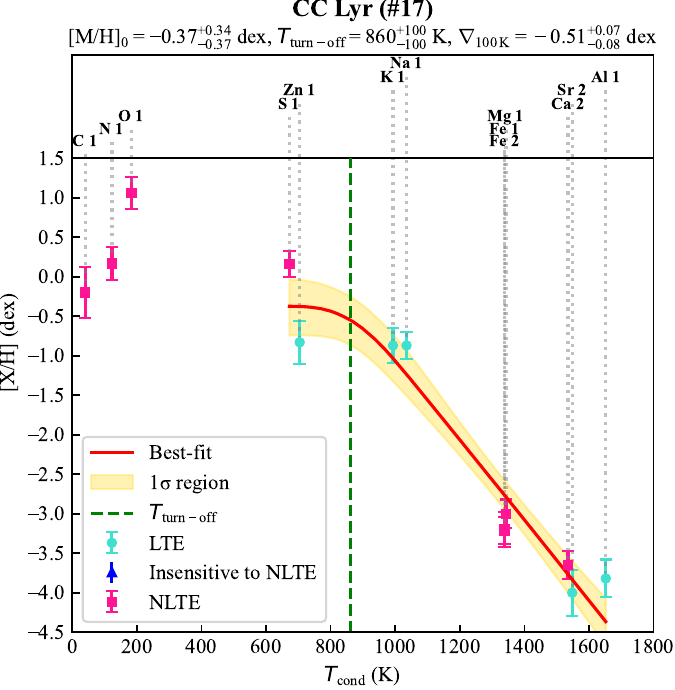}
    \includegraphics[width=.405\linewidth]{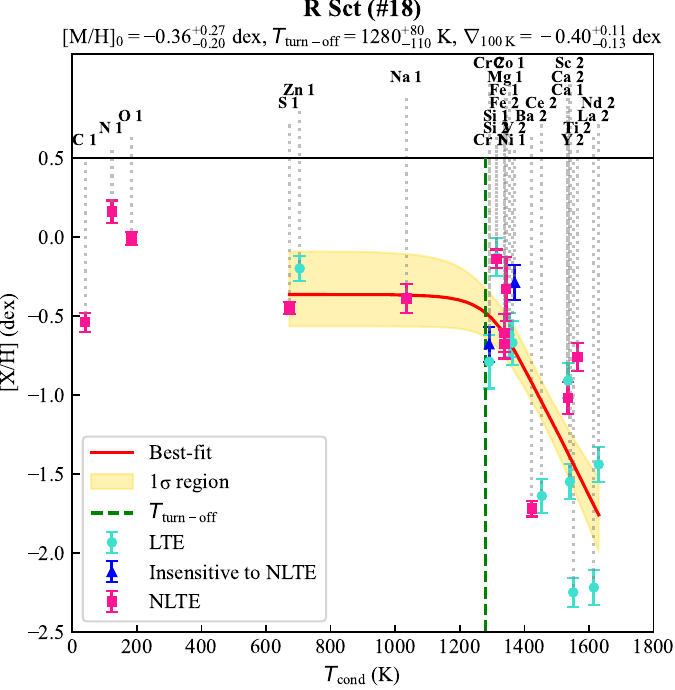}
    \includegraphics[width=.405\linewidth]{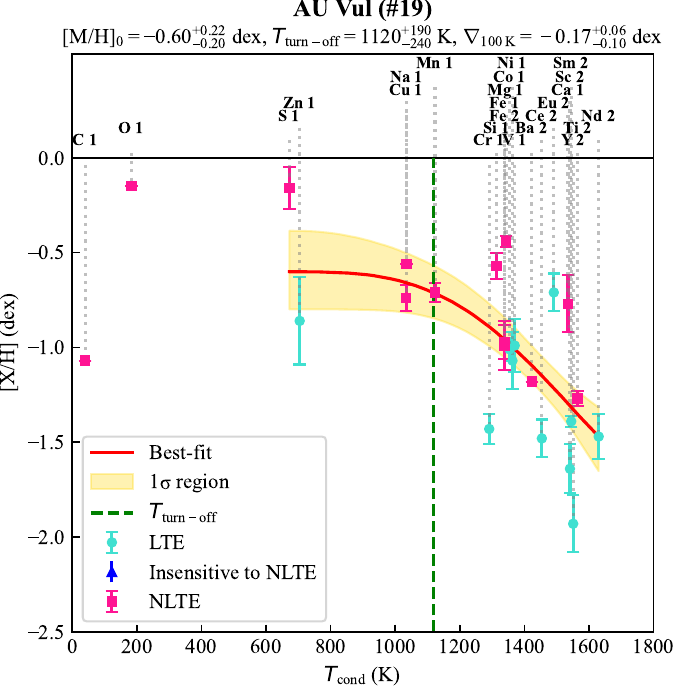}
    \includegraphics[width=.405\linewidth]{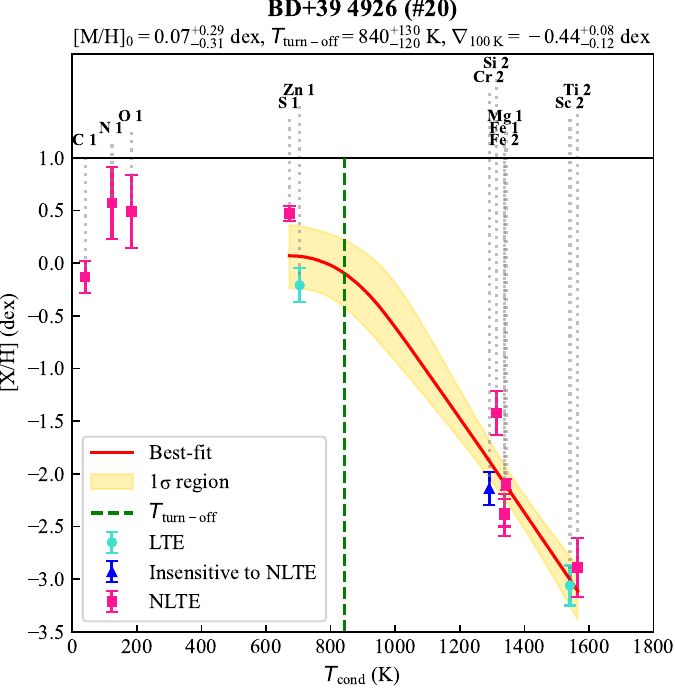}
    \caption[Elemental abundances of a subsample of post-AGB/post-RGB binaries with faint discs as functions of condensation temperature]{Elemental abundances of a subsample of post-AGB/post-RGB binaries with faint discs as functions of condensation temperature \citep{lodders2003CondensationTemperatures, wood2019CondensationTemperatures}. For ID explanation, see Table~\ref{tab:alllum}. The legend for the symbols and colours used is included within the plot. ``NLTE insensitive'' abundances are derived from spectral lines of \ion{V}{ii} and \ion{Cr}{ii}; for more details, see Section~\ref{ssec:anaspc}).}\label{fig:dpl1}
\end{figure*}

\begin{figure}
    \centering
    \includegraphics[width=.8\linewidth]{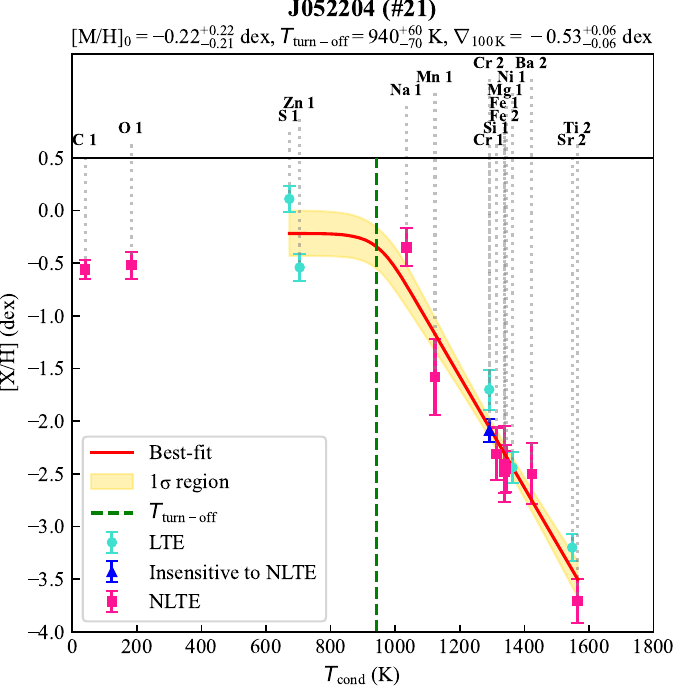}
    \includegraphics[width=.8\linewidth]{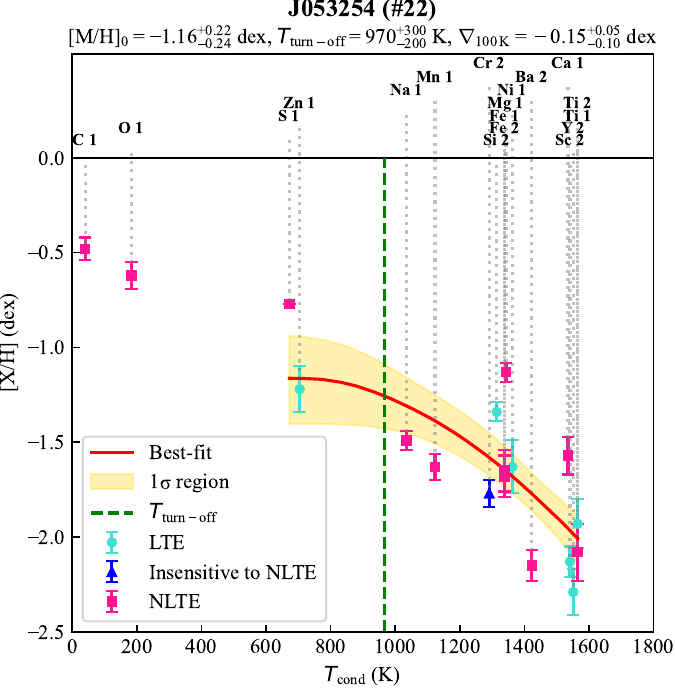}
    \caption[Elemental abundances of a subsample of post-AGB/post-RGB binaries with faint discs as functions of condensation temperature]{Elemental abundances of a subsample of post-AGB/post-RGB binaries with faint discs as functions of condensation temperature \citep{lodders2003CondensationTemperatures, wood2019CondensationTemperatures}. For ID explanation, see Table~\ref{tab:alllum}. The legend for the symbols and colours used is included within the plot. ``NLTE insensitive'' abundances are derived from spectral lines of \ion{V}{ii} and \ion{Cr}{ii}; for more details, see Section~\ref{ssec:anaspc}).}\label{fig:dpl2}
\end{figure}

In Table~\ref{tab:respar}, we present the derived atmospheric parameters and selected elemental abundance ratios of faint disc targets ([Zn/Ti], [Zn/Fe], [S/Ti], and C/O). In Figures~\ref{fig:dpl1} and \ref{fig:dpl2}, we mark the derived LTE abundances with cyan circles. In \ref{app:abu}, we present the derived LTE [X/H] abundances of faint disc targets.

\subsection{Calculation of NLTE corrections using \texttt{pySME}}\label{ssec:ananlt}
Departures from LTE generally (but not always) increase with higher $T_{\rm eff}$ and decreasing [Fe/H], particularly for weak lines of neutral minority species, including \ion{Fe}{i} and \ion{Ti}{i}, because of the stronger UV radiation field that can drive overionisation \citep[e.g.,][]{lind2024NLTEreview}. The departures also tend to increase with lower $\log g$ owing to less efficient collisional processes that would bring the system closer to LTE. As such, NLTE effects are important to consider when characterising most post-AGB/post-RGB stars ($T_{\rm eff}\gtrsim6\,000$\,K, $\log g\lesssim1$\,dex, [Fe/H]$\lesssim-1.5$\,dex).

In this study, we expand our NLTE analysis beyond that of \citet{mohorian2025TransitionDiscs}. Previously, in \citet{mohorian2025TransitionDiscs}, we used \texttt{Balder} \citep{amarsi2018Balder} to calculate NLTE corrections for individual spectral lines of C, N, O, Na, Mg, Al, Si, S, K, Ca, and Fe by matching the LTE and NLTE equivalent widths. In this work, we performed a more extensive analysis by adding Ti, Mn, Cu, and Ba to the list of NLTE-corrected elements. This scaling-up was motivated by the need to enhance the coverage of the depletion profiles for our targets, and therefore demanded a more efficient computational approach. We opted to use the 1D \texttt{pySME} code \citep{wehrhahn2023pySME}, which has the functionality to generate NLTE spectra with the help of grids of departure coefficients $\beta_k = \frac{n_k^{\rm NLTE}}{n_k^{\rm LTE}}$ \citep[pre-computed in \texttt{Balder} for each energy level $k$ with population $n_k$;][]{amarsi2020NLTEgalah}. Since the departure coefficients used in \texttt{pySME} were calculated for MARCS model atmospheres \citep{gustafsson2008MARCS}, which are limited for hot metal-poor giants, we set $T_{\rm eff}\,=\,5\,750$\,K and/or $\log g\,=\,0.5$\,dex in NLTE calculations for all targets with higher $T_{\rm eff}$ and lower $\log g$, respectively. In Table~\ref{tab:nltmod}, we list the references for the model atoms used to calculate the departure coefficient grids. The NLTE corrections were determined as the differences between the NLTE and LTE abundances, as explained below.

To calculate NLTE abundances, we synthesised the LTE spectral lines in \texttt{pySME} using observed equivalent widths and derived LTE atmospheric parameters. Then, we fitted these lines with NLTE spectral lines. The differences between NLTE and LTE abundances derived from the \texttt{pySME} fitted lines provided the relative abundance correction $\Delta_i^{\rm diff.}=[{\rm X/H}]_i^{\rm NLTE} - [{\rm X/H}]_i^{\rm LTE}$ for each studied spectral line $i$ (for average values of these corrections, see Section~\ref{ssec:dscall}). To calculate the uncertainty of NLTE abundance for an ionisation of an element, we added in quadratures the systematic uncertainty of LTE abundance (abundance deviations within the error bars of atmospheric parameters) and random uncertainty of individual NLTE abundances (0.1 dex for single line, standard deviation for more lines).

In \ref{app:abu}, we provide the derived NLTE [X/H] abundances of faint disc targets. In Figures~\ref{fig:dpl1} and \ref{fig:dpl2}, we present `NLTE-insensitive' abundances (\ion{V}{ii} and \ion{Cr}{ii}; navy triangles) and NLTE-corrected abundances (pink squares) of faint disc targets. In \ref{app:dpl}, we show that NLTE corrections have a minor impact on general trends in depletion profiles, although these corrections are essential for the derivation of precise individual abundances.

\begin{table}[ht]
    \centering
    \footnotesize
    \caption{The references for model atoms used in this study (see Section~\ref{ssec:ananlt}).} \label{tab:nltmod}
    \begin{tabular}{|c|c|}
    \hline
        Element & Reference \\ \hline
        C & \citet{amarsi2020NLTEgalah} \\
        N & \citet{amarsi2020NLTEgalah} \\
        O & \citet{amarsi2020NLTEgalah} \\
        Na & \citet{amarsi2020NLTEgalah} \\
        Mg & \citet{amarsi2020NLTEgalah} \\
        Al & \citet{amarsi2020NLTEgalah} \\
        Si & \citet{amarsi2020NLTEgalah} \\
        S & \citet{amarsi2025Sulphur} \\
        K & \citet{amarsi2020NLTEgalah} \\
        Ca & \citet{amarsi2020NLTEgalah} \\
        Ti & \citet{mallinson2024NLTE} \\
        Mn & \citet{amarsi2020NLTEgalah} \\
        Fe & \citet{amarsi2022NLTEiron} \\
        Cu & \citet{calsikan2025CuNLTE} \\
        Ba & \citet{amarsi2020NLTEgalah} \\ \hline
    \end{tabular}\\
    \vspace{-0.5cm}
\end{table}

\subsection{Quantification of depletion efficiency using three free parameters}\label{ssec:anares}
Depletion efficiency is measured primarily using volatile-to-refractory abundance ratios, including [Zn/Ti], [Zn/Fe], and [S/Ti] ratios (see Section~\ref{sec:int}). Although these abundance ratios allow for comparison with the literature, they do not fully determine which elements are significantly affected by depletion. Another depletion parameter, $T_{\rm turn-off}$ was traditionally determined by visual inspection, making it subjective and difficult to apply consistently. Furthermore, the high-temperature end of the depletion profile used to be categorised as `saturated' or `plateau' pattern (see Section~\ref{sec:int}). Although this pattern classification is intuitive, it lacks a quantitative basis and does not allow direct comparisons between different systems.

In Table~\ref{tab:respar}, we present [Zn/Ti], [Zn/Fe], and [S/Ti] ratios of faint disc targets (see Section~\ref{ssec:anaspc}). We note that we used NLTE-corrected abundances of S, Ti, and Fe to derive these volatile-to-refractory abundance ratios (see Section~\ref{ssec:ananlt}). In Table~\ref{tab:litpar}, we provide the reported $T_{\rm turn-off}$ values for faint disc targets from literature studies. In Table~\ref{tab:litpar}, we list the reported patterns of depletion profiles for faint disc targets from literature studies.

In this study, we investigated the efficiency of depletion in post-AGB/post-RGB binaries by exploring the shape of the depletion profiles (see Figures~\ref{fig:dpl1} and \ref{fig:dpl2}). To achieve this goal, we introduce a new approach to quantify depletion efficiency by fitting depletion profiles with two-piece linear functions, which have three free parameters:
\begin{itemize}
    \item The break of the two-piece linear function separates weakly and significantly depleted elements (elements with lower and higher condensation temperatures, respectively; see Section~\ref{sec:int}) and is represented by the turn-off temperature $T_{\rm turn-off}$.
    \item The left linear piece has a zero slope and describes the abundances of weakly depleted elements with $T_{\rm cond}<T_{\rm turn-off}$ ($\text{[X/H]} = \text{[M/H]}_0$, where [M/H]$_0$ is the initial metallicity).
    \item The right linear piece describes the abundances of significantly depleted elements with $T_{\rm cond}\geq T_{\rm turn-off}$ ($\text{[X/H]} = \text{[M/H]}_0+\nabla_{\rm 100\,K}\cdot\frac{T_{\rm cond}-T_{\rm turn-off}}{\rm 100\,K}$, where $\nabla_{\rm 100\,K}$ is the depletion scale in dex per 100\,K).
\end{itemize}

The observed depletion profiles were fitted using a Bayesian approach implemented in PyMC5, using the No-U-Turn Sampler (NUTS) with 4 chains of 10\,000 tuning and 10\,000 sampling iterations each. The priors were placed on the model parameters as follows: initial metallicity $\text{[M/H]}_0$ and depletion scale $\nabla_{\rm 100\,K}$ were assigned normal distributions centered at 0 dex with a standard deviation of 10 dex, while turn-off temperature $T_{\rm turn-off}$ was given a uniform distribution spanning the full temperature range of each individual depletion profile (excluding CNO elements; see Section~\ref{sec:int}). Parameter uncertainties were modelled using half-normal distributions with a mean of 0 dex and a standard deviation of 10 dex. These weakly informative priors were chosen to reflect broad physical plausibility without enforcing strong constraints. Elemental abundances were treated as observational data with equal weighting across all fitted elements. Posterior predictions were obtained by sampling from the posterior distributions, and credible intervals were derived from the $16^{\rm th}$ and $84^{\rm th}$ percentiles.

Together, [M/H]$_0$, $T_{\rm turn-off}$, and $\nabla_{\rm 100\,K}$ provide a simplified, yet comprehensive description of the observed depletion profiles in post-AGB and post-RGB binaries. In Figures~\ref{fig:dpl1} and \ref{fig:dpl2}, we present the depletion profile fits for faint disc targets (red solid lines) with corresponding $1\sigma$-areas (yellow shaded regions). We note that CNO elements were excluded from the depletion fitting, as CNO surface abundances were significantly altered by nucleosynthetic and mixing processes during the AGB/RGB evolution, on scales comparable to and indiscernible from those of the depletion process (see Section~\ref{sec:int}).

We found that all derived NLTE depletion profiles of the faint disc targets show saturation (see Section~\ref{sec:int}). In Table~\ref{tab:alllum}, we list the adopted luminosities, atmospheric parameters, and depletion parameters of the faint disc targets. To further investigate depletion in faint disc targets, we examined how the derived depletion parameters vary with effective temperature, adopted luminosity, orbital period, and IR/stellar luminosity, as these parameters were shown to be moderately correlating with depletion in transition disc targets \citep{mohorian2025TransitionDiscs}. Although we found a wide range of depletion parameters among faint disc targets, the small sample size prevented detection of any significant correlations or trends (see the distribution of faint disc targets in Figures~\ref{fig:dplteff}, \ref{fig:dpllum}, \ref{fig:dplporb}, and \ref{fig:dpllir}). To build on these results, we extend our analysis to include full and transition disc targets, enabling a comparison of depletion profiles across different disc types.

\begin{table*}[!ht]
    \centering
    \resizebox{\columnwidth}{!}{
    \caption[Adopted luminosities, atmospheric parameters, and depletion parameters (see Section~\ref{sec:ana}) of the post-AGB/post-RGB binaries with faint, full, and transition discs (see Section~\ref{sec:dsc})]{Adopted luminosities and homogeneously derived atmospheric and depletion parameters (using \texttt{E-iSpec} and \texttt{pySME}; see Section~\ref{sec:ana}) of the post-AGB/post-RGB binaries with faint, full, and transition discs. For more details, see Section~\ref{sec:dsc}.} \label{tab:alllum}
    \begin{tabular}{|c|c|c|c|c|c|c|c|c|c|}
    \hline
        \textbf{ID} & \textbf{Name} & \boldmath$\log\frac{L_{\rm adopted}}{L_\odot}$ & \boldmath$T_{\rm eff}$\,\textbf{(K)} & \boldmath$\log g$\,\textbf{(dex)} & \textbf{[Fe/H] (dex)} & \boldmath$\xi_{\rm t}$\,\textbf{(km/s)} & \textbf{[M/H]}\boldmath$_0$ \textbf{(dex)} & \boldmath$T_{\rm turn-off}$ \textbf{(K)} & \boldmath$\nabla_{\rm 100\,K}$ \textbf{(dex)} \\ \hline
        \multicolumn{10}{|c|}{\textit{Full disc targets \citep{mohorian2024EiSpec}}} \\ \hline
        1 & SZ Mon & $2.58\,\pm\,0.10$ & $5460\,\pm\,60$ & $0.93\,\pm\,0.10$ & $-0.50\,\pm\,0.05$ & $4.37\,\pm\,0.08$ & $-0.32_{-0.11}^{+0.13}$ & $1270_{-90}^{+70}$ & $-0.25_{-0.07}^{+0.06}$ \\
        2 & DF Cyg & $2.82\,\pm\,0.07$ & $5770\,\pm\,70$ & $1.92\,\pm\,0.09$ & $+0.05\,\pm\,0.05$ & $3.97\,\pm\,0.03$ & $+0.25_{-0.12}^{+0.14}$ & $1280_{-80}^{+80}$ & $-0.25_{-0.08}^{+0.06}$ \\ \hline
        \multicolumn{10}{|c|}{\textit{Transition disc targets \citep{mohorian2025TransitionDiscs}}} \\ \hline
        3 & CT Ori & $3.26\,\pm\,0.16$ & $5940\,\pm\,120$ & $1.01\,\pm\,0.18$ & $-1.89\,\pm\,0.11$ & $3.37\,\pm\,0.10$ & $-0.49_{-0.25}^{+0.24}$ & $840_{-110}^{+140}$ & $-0.24_{-0.04}^{+0.04}$ \\
        4 & ST Pup & $2.96\,\pm\,0.17$ & $5340\,\pm\,80$ & $0.20\,\pm\,0.10$ & $-1.92\,\pm\,0.08$ & $2.83\,\pm\,0.03$ & $-0.67_{-0.19}^{+0.19}$ & $840_{-100}^{+100}$ & $-0.24_{-0.03}^{+0.03}$ \\
        5 & RU Cen & $3.50\,\pm\,0.27$ & $6120\,\pm\,80$ & $1.46\,\pm\,0.15$ & $-1.93\,\pm\,0.08$ & $3.26\,\pm\,0.10$ & $-0.93_{-0.14}^{+0.13}$ & $750_{-60}^{+100}$ & $-0.14_{-0.02}^{+0.02}$ \\
        6 & AC Her & $3.71\,\pm\,0.21$ & $6140\,\pm\,100$ & $1.27\,\pm\,0.16$ & $-1.47\,\pm\,0.08$ & $3.92\,\pm\,0.12$ & $-0.67_{-0.13}^{+0.12}$ & $850_{-110}^{+120}$ & $-0.15_{-0.02}^{+0.02}$ \\
        7 & AD Aql & $2.84\,\pm\,0.29$ & $6200\,\pm\,170$ & $1.67\,\pm\,0.45$ & $-2.20\,\pm\,0.09$ & $2.98\,\pm\,0.36$ & $-0.07_{-0.21}^{+0.20}$ & $800_{-80}^{+80}$ & $-0.38_{-0.05}^{+0.04}$ \\
        8 & EP Lyr & $3.96\,\pm\,0.23$ & $6270\,\pm\,160$ & $1.24\,\pm\,0.18$ & $-2.03\,\pm\,0.17$ & $2.48\,\pm\,0.10$ & $-0.65_{-0.16}^{+0.16}$ & $730_{-40}^{+80}$ & $-0.20_{-0.03}^{+0.03}$ \\
        9 & DY Ori & $3.50\,\pm\,0.17$ & $6160\,\pm\,70$ & $0.88\,\pm\,0.14$ & $-2.03\,\pm\,0.04$ & $2.48\,\pm\,0.09$ & $+0.20_{-0.33}^{+0.34}$ & $890_{-130}^{+120}$ & $-0.38_{-0.08}^{+0.07}$ \\
        10 & AF Crt & $2.51\,\pm\,0.17$ & $6110\,\pm\,110$ & $0.96\,\pm\,0.21$ & $-2.47\,\pm\,0.05$ & $4.87\,\pm\,0.16$ & $-0.34_{-0.37}^{+0.36}$ & $890_{-130}^{+140}$ & $-0.35_{-0.08}^{+0.07}$ \\
        11 & GZ Nor & $2.71\,\pm\,0.29$ & $5100\,\pm\,80$ & $0.49\,\pm\,0.16$ & $-1.65\,\pm\,0.07$ & $4.33\,\pm\,0.05$ & $-0.89_{-0.15}^{+0.14}$ & $750_{-50}^{+110}$ & $-0.11_{-0.02}^{+0.02}$ \\
        12 & V1504 Sco & $3.71\,\pm\,0.35$ & $5980\,\pm\,90$ & $0.98\,\pm\,0.17$ & $-1.05\,\pm\,0.07$ & $4.29\,\pm\,0.05$ & $+0.14_{-0.17}^{+0.19}$ & $950_{-100}^{+70}$ & $-0.31_{-0.05}^{+0.05}$ \\
        13 & J050304 & $3.46\,\pm\,0.20$ & $5750\,\pm\,100$ & $0.00\,\pm\,0.18$ & $-2.61\,\pm\,0.05$ & $2.20\,\pm\,0.01$ & $-0.77_{-0.34}^{+0.33}$ & $840_{-110}^{+120}$ & $-0.34_{-0.07}^{+0.06}$ \\
        14 & J053150 & $3.99\,\pm\,0.19$ & $6160\,\pm\,130$ & $1.38\,\pm\,0.20$ & $-2.48\,\pm\,0.04$ & $4.28\,\pm\,0.12$ & $-0.08_{-0.34}^{+0.39}$ & $960_{-120}^{+80}$ & $-0.42_{-0.08}^{+0.08}$ \\ \hline
        \multicolumn{10}{|c|}{\textit{Faint disc targets (this study)}} \\ \hline
        15 & SS Gem & $3.40\,\pm\,0.20$ & $6500\,\pm\,90$ & $1.96\,\pm\,0.11$ & $-1.02\,\pm\,0.12$ & $4.08\,\pm\,0.09$ & $-0.33_{-0.10}^{+0.12}$ & $1030_{-100}^{+80}$ & $-0.17_{-0.03}^{+0.03}$ \\
        16 & V382 Aur & $3.35\,\pm\,0.15$ & $6020\,\pm\,150$ & $0.38\,\pm\,0.27$ & $-1.82\,\pm\,0.08$ & $4.14\,\pm\,0.13$ & $-0.92_{-0.24}^{+0.22}$ & $820_{-110}^{+170}$ & $-0.14_{-0.04}^{+0.04}$ \\
        17 & CC Lyr & $2.99\,\pm\,0.16$ & $6470\,\pm\,250$ & $1.64\,\pm\,0.50$ & $-3.80\,\pm\,0.03$ & $4.14\,\pm\,1.00$ & $-0.38_{-0.36}^{+0.34}$ & $860_{-100}^{+100}$ & $-0.51_{-0.08}^{+0.07}$ \\
        18 & R Sct & $3.10\,\pm\,0.25$ & $5630\,\pm\,70$ & $0.93\,\pm\,0.11$ & $-0.60\,\pm\,0.11$ & $2.93\,\pm\,0.03$ & $-0.37_{-0.20}^{+0.26}$ & $1280_{-110}^{+80}$ & $-0.40_{-0.12}^{+0.11}$ \\
        19 & AU Vul & $3.82\,\pm\,0.20$ & $4740\,\pm\,40$ & $0.00\,\pm\,0.09$ & $-0.95\,\pm\,0.08$ & $4.32\,\pm\,0.04$ & $-0.61_{-0.20}^{+0.22}$ & $1130_{-240}^{+180}$ & $-0.17_{-0.10}^{+0.06}$ \\
        20 & BD+39 4926 & $3.79\,\pm\,0.14$ & $7470\,\pm\,40$ & $0.52\,\pm\,0.05$ & $-2.37\,\pm\,0.18$ & $2.21\,\pm\,0.25$ & $+0.08_{-0.31}^{+0.29}$ & $850_{-110}^{+130}$ & $-0.44_{-0.12}^{+0.08}$ \\
        21 & J052204 & $3.31\,\pm\,0.11$ & $7020\,\pm\,100$ & $1.65\,\pm\,0.11$ & $-2.49\,\pm\,0.10$ & $2.36\,\pm\,0.04$ & $-0.22_{-0.21}^{+0.21}$ & $940_{-70}^{+60}$ & $-0.53_{-0.06}^{+0.06}$ \\
        22 & J053254 & $3.84\,\pm\,0.28$ & $6040\,\pm\,90$ & $0.47\,\pm\,0.13$ & $-1.66\,\pm\,0.09$ & $2.36\,\pm\,0.03$ & $-1.17_{-0.24}^{+0.23}$ & $970_{-200}^{+300}$ & $-0.15_{-0.10}^{+0.05}$ \\ \hline
        %23 & BD+28 772 & $3.50\,\pm\,0.48$ & $6880\,\pm\,100$ & $0.96\,\pm\,0.16$ & $-0.42\,\pm\,0.10$ & $2.14\,\pm\,0.03$ & $-0.32_{-0.04}^{+0.05}$ & $1280_{-260}^{+160}$ & $-0.04_{-0.03}^{+0.02}$ \\ \hline
    \end{tabular}
    }\\
    \textbf{Notes:} Post-AGB and post-RGB binaries are separated by the luminosity ($\log(L/L_\odot)\,\gtrsim\,3.4$ and $\log(L/L_\odot)\,\lesssim\,3.4$, respectively). Atmospheric parameters of GZ Nor (\#11) were revised using a more conservative line list to decrease the line-to-line scatter of elemental abundances (see Table~\ref{tabA:linlst}).
    \vspace{-0.5cm}
\end{table*}

\section{Discussion}\label{sec:dsc}
In this section, we examine the depletion characteristics of the faint disc targets and explore their broader implications for disc evolution. In Section~\ref{ssec:dscall}, we compare faint disc systems to a previously analysed subset of post-AGB and post-RGB binaries hosting full and transition discs. In Section~\ref{ssec:dscdsp}, we discuss how our findings inform current understanding of disc dissipation processes in these evolved binaries.

\subsection{Depletion profiles across faint, full, and transition disc targets}\label{ssec:dscall}
To investigate how photospheric depletion varies across different evolutionary stages of post-AGB/post-RGB binaries, we extended our target sample by incorporating full and transition disc targets from \citet{mohorian2024EiSpec} and \citet{mohorian2025TransitionDiscs}. We hereafter refer to this curated set of targets as the {\it combined sample}. Although we adopted the published LTE abundances and luminosities for these targets, we recalculated their NLTE corrections using \texttt{pySME} (as noted in Section~\ref{ssec:ananlt}), replacing the original \texttt{Balder}-based corrections used in the earlier studies \citep{mohorian2025TransitionDiscs}. These revised corrections were then applied to the LTE abundances following the same procedure (see Section~\ref{ssec:ananlt}) used for the faint disc sample, ensuring a consistent framework across all targets. In \ref{app:dpl}, we present the full details of recalculated abundances, applied NLTE corrections, and comparisons with previously published values. In Table~\ref{tab:alllum}, we summarise the adopted luminosities, atmospheric parameters, and derived depletion parameters of the combined sample.

\begin{figure}[!ht]
    \centering
    \includegraphics[trim={1cm 1cm 0.8cm 1cm}, width=0.8\columnwidth]{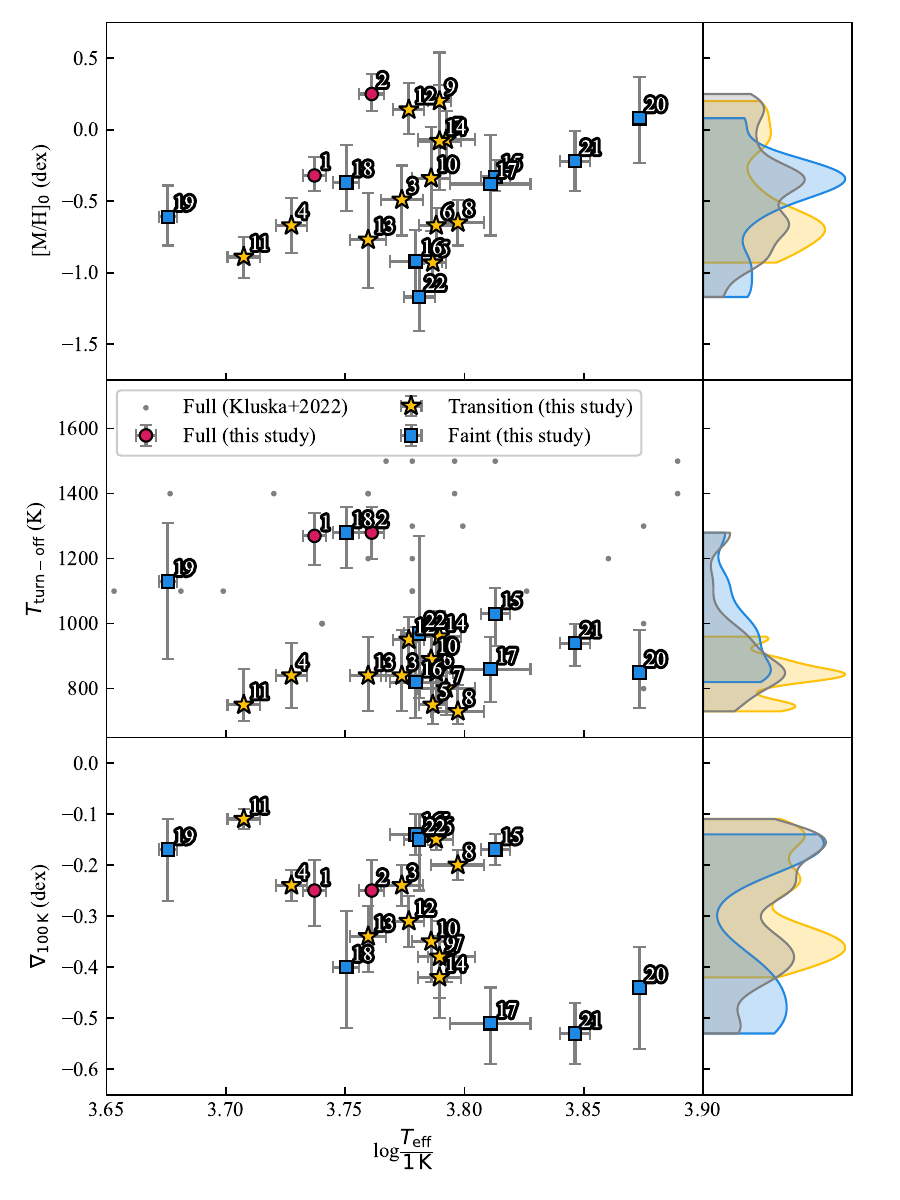}
    \caption[Distributions of depletion pattern parameters with effective temperature as a proxy for age on post-AGB/post-RGB track (\textit{upper panel}: initial metallicity {[M/H]}$_0$, \textit{middle panel}: $T_{\rm turn-off}$, \textit{lower panel}: depletion gradient $\nabla_{\rm100K}$)]{Distributions of depletion pattern parameters with effective temperature as a proxy for age on post-AGB/post-RGB track (\textit{upper panel}: initial metallicity [M/H]$_0$, \textit{middle panel}: $T_{\rm turn-off}$, \textit{lower panel}: depletion gradient $\nabla_{\rm100\,K}$). For target ID specification, see Table~\ref{tab:alllum}. Red circles represent full disc targets, yellow stars represent transition disc targets, blue squares represent faint disc targets. Grey dots in $T_{\rm turn-off}$ subplot represent the Galactic full disc targets with visually estimated values of $T_{\rm turn-off}$ \citep{kluska2022GalacticBinaries}. Small right panels display the distributions of depletion parameters within the targets hosting transition discs (yellow), faint discs (blue), and all types of discs (gray). For more details on depletion pattern parameters, see Section~\ref{sec:dsc}.}\label{fig:dplteff}
\end{figure}

\begin{figure}[!ht]
    \centering
    \includegraphics[trim={1cm 1cm 0.8cm 1cm}, width=0.8\columnwidth]{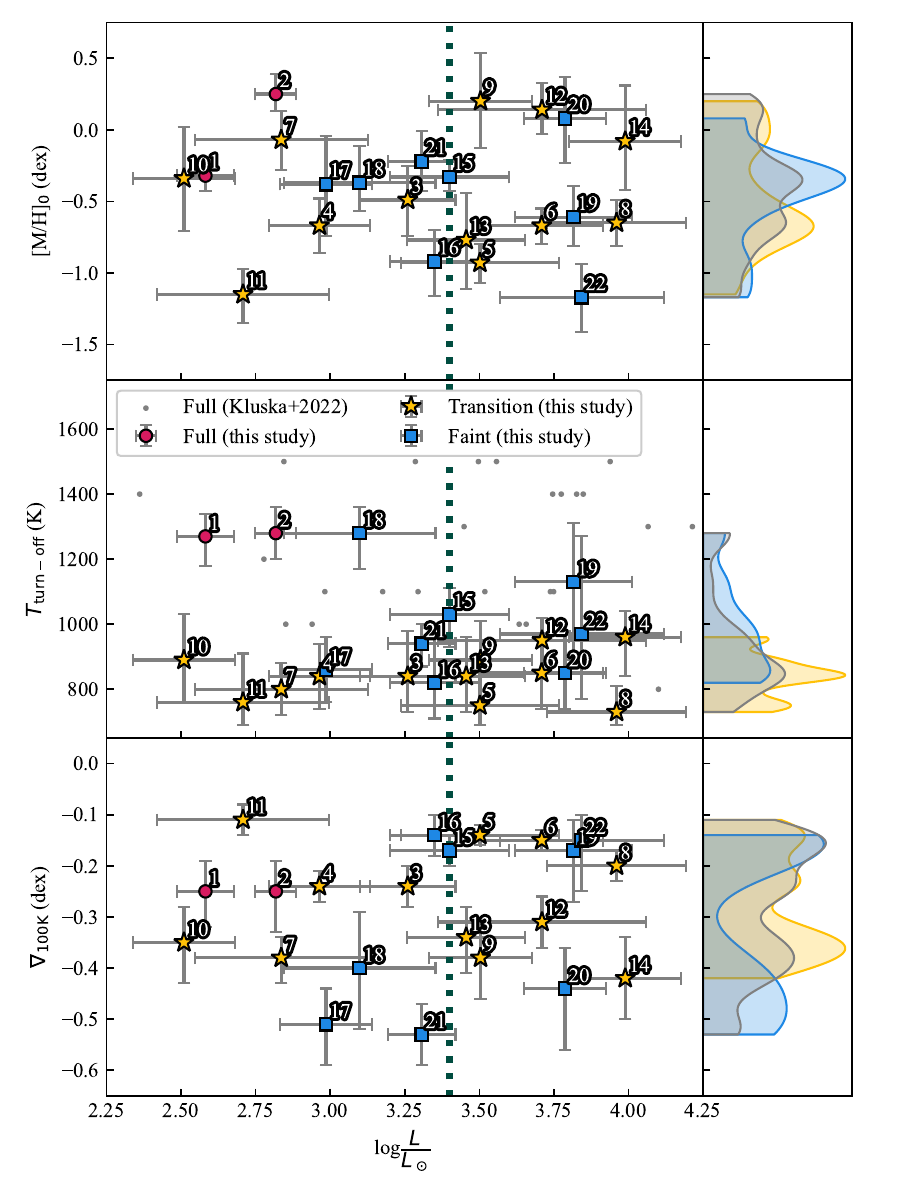}
    \caption[Distributions of depletion pattern parameters with adopted luminosity (\textit{upper panel}: initial metallicity {[M/H]}$_0$, \textit{middle panel}: $T_{\rm turn-off}$, \textit{lower panel}: depletion gradient $\nabla_{\rm100K}$)]{Distributions of depletion pattern parameters with adopted luminosity (\textit{upper panel}: initial metallicity [M/H]$_0$, \textit{middle panel}: $T_{\rm turn-off}$, \textit{lower panel}: depletion gradient $\nabla_{\rm100\,K}$). For target ID specification, see Table~\ref{tab:alllum}. Red circles represent full disc targets, yellow stars represent transition disc targets, blue squares represent faint disc targets. Vertical dotted line represents rough demarcation between post-AGB and post-RGB binaries ($L\,\sim\,2\,500\,L_\odot$; $\log\frac{L}{L_\odot}\,\sim\,3.4$). Grey dots in $T_{\rm turn-off}$ subplot represent the Galactic full disc targets with visually estimated values of $T_{\rm turn-off}$ \citep{kluska2022GalacticBinaries}. Small right panels display the distributions of depletion parameters within the targets hosting transition discs (yellow), faint discs (blue), and all types of discs (gray). For more details on depletion pattern parameters, see Section~\ref{sec:dsc}.}\label{fig:dpllum}
\end{figure}

\begin{figure}[!ht]
    \centering
    \includegraphics[trim={1cm 1cm 0.8cm 1cm}, width=0.8\columnwidth]{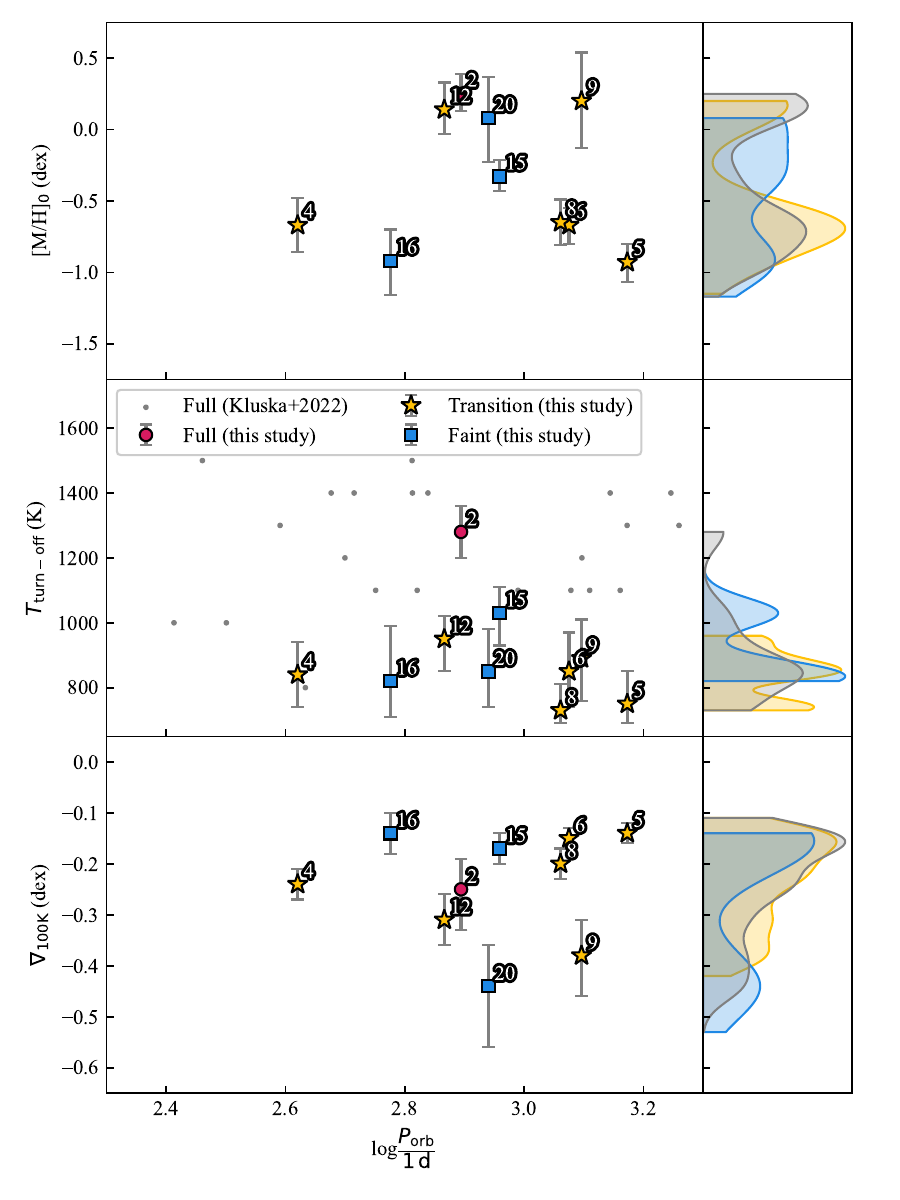}
    \caption[Distributions of depletion pattern parameters with orbital period (\textit{upper panel}: initial metallicity {[M/H]}$_0$, \textit{middle panel}: $T_{\rm turn-off}$, \textit{lower panel}: depletion gradient $\nabla_{\rm100K}$)]{Distributions of depletion pattern parameters with orbital period (\textit{upper panel}: initial metallicity [M/H]$_0$, \textit{middle panel}: $T_{\rm turn-off}$, \textit{lower panel}: depletion gradient $\nabla_{\rm100\,K}$). For target ID specification, see Table~\ref{tab:alllum}. Red circles represent full disc targets, yellow stars represent transition disc targets, blue squares represent faint disc targets. Grey dots in $T_{\rm turn-off}$ subplot represent the Galactic full disc targets with visually estimated values of $T_{\rm turn-off}$ \citep{kluska2022GalacticBinaries}. Small right panels display the distributions of depletion parameters within the targets hosting transition discs (yellow), faint discs (blue), and all types of discs (gray). For more details on depletion pattern parameters, see Section~\ref{sec:dsc}.}\label{fig:dplporb}
\end{figure}

\begin{figure}[!ht]
    \centering
    \includegraphics[trim={1cm 1cm 0.8cm 1cm}, width=0.8\columnwidth]{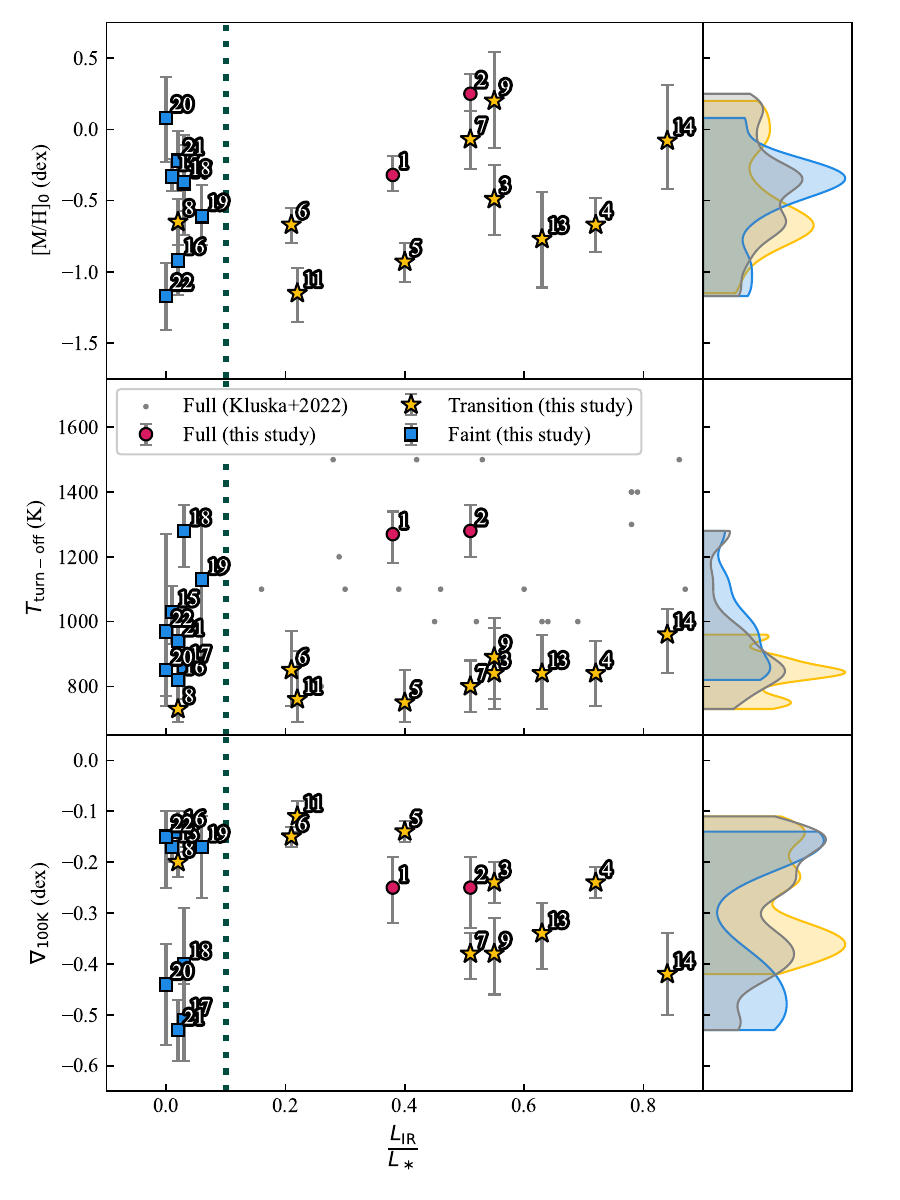}
    \caption[Distributions of depletion pattern parameters with IR/stellar luminosity ratio (\textit{upper panel}: initial metallicity {[M/H]}$_0$, \textit{middle panel}: $T_{\rm turn-off}$, \textit{lower panel}: depletion gradient $\nabla_{\rm100K}$)]{Distributions of depletion pattern parameters with IR/stellar luminosity ratio (\textit{upper panel}: initial metallicity [M/H]$_0$, \textit{middle panel}: $T_{\rm turn-off}$, \textit{lower panel}: depletion gradient $\nabla_{\rm100\,K}$). Targets with $L_{\rm IR}/L_\ast>0.9$ are excluded from the figure because these systems are viewed at high inclinations (i.e., close to edge-on). In such cases, the observed IR excess is strongly affected by geometric effects, requiring a more careful and detailed interpretation before meaningful conclusions can be drawn. For target ID specification, see Table~\ref{tab:alllum}. Red circles represent full disc targets, yellow stars represent transition disc targets, blue squares represent faint disc targets. Vertical dotted line represents rough demarcation between faint and full/transition disc targets ($L_{\rm IR}/L_\ast\,=\,0.1$). Grey dots in $T_{\rm turn-off}$ subplot represent the Galactic full disc targets with visually estimated values of $T_{\rm turn-off}$ \citep{kluska2022GalacticBinaries}. Small right panels display the distributions of depletion parameters within the targets hosting transition discs (yellow), faint discs (blue), and all types of discs (gray). For more details on depletion pattern parameters, see Section~\ref{sec:dsc}.}\label{fig:dpllir}
\end{figure}

To assess depletion efficiency across post-AGB/post-RGB binaries, we analysed how the depletion parameters (initial metallicity [M/H]$_0$, turn-off temperature $T_{\rm turn-off}$, and depletion scale $\nabla_{\rm 100\,K}$; see Section~\ref{ssec:anares}) vary across the combined sample of faint, full, and transition disc targets. Specifically, we explored trends with stellar effective temperature, bolometric luminosity, orbital period, and IR/stellar luminosity ratio (see Figures~\ref{fig:dplteff}, \ref{fig:dpllum}, \ref{fig:dplporb}, and \ref{fig:dpllir}). These four parameters were selected based on the following considerations:
\begin{itemize}
    \item Effective temperature ($T_{\rm eff}$) can serve as a proxy for evolutionary stage along the post-AGB/post-RGB track \citep{oomen2019depletion, martin2025ModellingDepletion}. Although $T_{\rm eff}$ is not a perfect indicator -- since accretion is known to stall the evolution of post-AGB/post-RGB stars -- it remains informative. \citet{martin2025ModellingDepletion} showed that with accretion rates up to $10^{-6}\,M_\odot$/year and disc masses up to $3\cdot10^{-2}\,M_\odot$, the post-AGB/post-RGB phase may be extended by up to a factor of 10. Despite this, $T_{\rm eff}$ still provides a reasonable baseline for assessing whether a target is in an earlier or later stage of post-AGB/post-RGB evolution.
    \item Bolometric luminosity offers a means of distinguishing between more luminous post-AGB stars and less luminous post-RGB stars, with an approximate boundary at $L\,\sim\,2\,500\,L_\odot$ ($\log\frac{L}{L_\odot}\,\sim\,3.4$), depending on the metallicity \citep{kamath2016PostRGBDiscovery}.
    \item Orbital period is an important tracer of binary interaction history \citep{vanwinckel2003Review, vanwinckel2018Binaries, oomen2020MESAdepletion}. However, this parameter is only available for only 10 of the 18 Galactic targets in our sample, limiting its statistical robustness and making it susceptible to small-number statistics and observational biases.
    \item The IR/stellar luminosity ratio ($L_{\rm IR}/L_\ast$) serves as a proxy for the disc mass or optical thickness \citep{deruyter2006discs}.  However, this quantity is affected by disc geometry. Disentangling this effect requires radiative transfer modelling constrained by spatially resolved observations, which lies beyond the scope of this study.
\end{itemize}

The main findings from our comparison of the NLTE-corrected depletion profiles across the combined sample are summarised below:
\begin{itemize}
    \item The initial metallicity [M/H]$_0$ shows significant target-to-target variability with no overarching trend, suggesting that the depletion process operates independently of initial metallicity.
    \item The $T_{\rm turn-off}$ values differ between disc types: both full disc post-RGB systems (SZ Mon, \#1; and DF Cyg, \#2) show $T_{\rm turn-off}\sim$1300\,K, while transition disc targets (both post-RGB and post-AGB) exhibit $T_{\rm turn-off}\sim$900\,K (see Figures~\ref{fig:dplteff}, \ref{fig:dpllum}, \ref{fig:dplporb}, and \ref{fig:dpllir}). This morphological difference suggests distinct disc-formation mechanisms. Based on this, we classified faint disc targets into two subgroups: i) full-like disc targets with $T_{\rm turn-off}\,>\,1\,100$\,K (R Sct, \#18; and AU Vul, \#19) and ii) transition-like disc targets with $T_{\rm turn-off}\,<\,1\,100$\,K (SS Gem, \#15; V382 Aur, \#16; CC Lyr, \#17; BD+39\,4926, \#20; J052204, \#21; and J053254, \#22). In future studies, we will explore this classification further through homogeneous chemical analysis of full disc targets in the Galaxy and in the LMC.
    \item The depletion scale $\nabla_{\rm 100\,K}$ shows a bimodal distribution among faint disc targets and a more stochastic distribution across the combined sample. Three high-$T_{\rm eff}$ faint disc targets (AU\,Vul, \#19; BD+39 4926, \#20; J052204, \#21) contribute strongly to the bimodality, showing steep depletion gradients ($\nabla_{\rm 100\,K} < -0.4$\,dex; see lower right panel of Figures~\ref{fig:dplteff}, \ref{fig:dpllum}, \ref{fig:dplporb}, and \ref{fig:dpllir}). The overall stochasticity of $\nabla_{\rm 100\,K}$ is consistent with the established picture of chemical depletion, in which refractory elements are rapidly sequestered into dust and prevented from re-accretion, potentially erasing earlier RGB/AGB nucleosynthetic signatures.
    \item Depletion is notably more efficient in post-AGB/post-RGB binaries than in Sun-like planet hosts. The average depletion scale is $\nabla_{\rm 100\,K}\,\sim\,-0.3$ dex in our sample, compared to $\nabla_{\rm 100\,K}\,\sim\,-0.01$ dex in young planet-hosting stars \citep{yun2024SolarDepletion}. This efficiency likely reflects more effective gas-dust fractionation in CBDs or smaller convective envelopes ($\sim\,10^{-3}\,-\,10^{-2}\,M_\odot$) that retain chemical imprints more easily \citep{oomen2020MESAdepletion}.
    \item Four LMC targets (J050304, \#13; J053150, \#14; J052204, \#21; J053254, \#22) follow the same depletion trends as Galactic systems, suggesting that initial metallicity has little influence on depletion efficiency.
    \item Transition disc target EP Lyr (\#8) with a wide inner gap \citep[based on mid-IR interferometry;][]{corporaal2023DiscParameters}, has a low IR/stellar luminosity ratio ($L_{\rm IR}/L_\ast = 0.02$) that belongs within the range for faint discs ($0<L_{\rm IR}/L_\ast<0.1$). This suggests a potential evolutionary connection between faint and transition disc systems.
    \item In Figure~\ref{fig:dpllir}, the $L_{\rm IR}/L_\ast$-$T_{\rm turn-off}$ subplot (middle panel) allows for segregation of full, transition, and faint disc targets. This further supports the idea that disc types might reflect different evolutionary stages. To validate this, we plan to expand our abundance analyses to full disc systems in the Galaxy and Magellanic Clouds \citep{kluska2022GalacticBinaries, kamath2014SMC, kamath2015LMC}.
    \item In Figure~\ref{fig:dpllir}, the $L_{\rm IR}/L_\ast$-$\nabla_{\rm 100\,K}$ subplot (lower panel) shows that transition disc targets with higher $L_{\rm IR}/L_\ast$ tend to exhibit steeper depletion slopes ($\nabla_{\rm 100\,K}$), with a statistically significant Spearman correlation ($\rho_s = -0.70$). However, this trend may be affected by system inclination and dust mass, which are difficult to disentangle without full orbital solutions (available for only 6 of 12 transition disc targets). As depletion scale is also shaped by disc evolution, further study is needed to interpret this correlation robustly.
\end{itemize}

Finally, although our results do not reveal a clear relation between stellar and depletion parameters, understanding the underlying physical and chemical mechanisms remains crucial. In particular, interpretation of the observed depletion parameter distributions requires developing detailed models of individual CBDs (including processes of dust condensation, inward transport, and gas-dust decoupling). To gain deeper insight into photospheric depletion observed in post-AGB/post-RGB binaries, our future work will focus on advancing the modelling of the depletion process through re-accretion, building on the initial works of \citet{oomen2019depletion, oomen2020MESAdepletion, martin2025ModellingDepletion}. This will involve incorporating more sophisticated disc physics, including photo-evaporation, binary-induced torques, irradiation from central binary, and spatial or temporal variations in the chemical composition of re-accreted material.


\subsection{Disc dissipation around post-AGB and post-RGB binaries}\label{ssec:dscdsp}
Post-AGB and post-RGB binary stars were traditionally identified by the significant IR excesses in their SEDs (see Section~\ref{sec:int}). However, faint disc targets with low IR excesses \citep[$\sim10\%$ of post-AGB/post-RGB binary sample in the Galaxy;][]{kluska2022GalacticBinaries} still show clear signs of photospheric depletion (see Table~\ref{tab:litpar}). This suggests that these faint disc systems likely hosted full or transition discs during an earlier phase of disc-binary interaction. Our findings support this interpretation and strengthen the categorisation of faint disc systems into full-like and transition-like subsets (see Section~\ref{ssec:dscall}).

Although the faint disc sample is not systematically hotter (i.e., more evolved; see Section~\ref{ssec:dscall}) than the full or transition disc samples, 7 of 8 faint disc targets -- SS Gem (\#15), V382 Aur (\#16), CC Lyr (\#17), R Sct (\#18), BD+39\,4926 (\#20), J052204 (\#21), and J053254 (\#22) -- have moderate-to-high effective temperatures ($T_{\rm eff}\,>\,5\,500$\,K). These targets might represent the progeny of full or transition disc targets that since dissipated, potentially through a combination of grain growth, radial drift, and photo-evaporation \citep[as proposed for young stars with dissipated protoplanetary discs; see][and references therein]{alexander2006Photoevaporation, alexander2014photoevaporation}. This hypothesis is plausible given reported jet-driven mass-loss rates of $10^{-8} - 10^{-6}\,M_\odot$/yr in post-AGB/post-RGB systems \citep{bollen2022Jets, verhamme2024DiscWindModelling}. Recently, \citet{deprins2024Jets} presented the first MHD models of such jets, reporting higher mass-loss rates ($10^{-6} - 10^{-3}\,M_\odot$/yr), though they noted these values are likely overestimated for post-AGB/post-RGB binaries. Even under conservative assumptions, a post-AGB/post-RGB system could lose $\sim\,10^{-3}\,M_\odot$ of CBD material over the $\sim\,10^5$-year duration of the post-AGB phase \citep{bertolami2016tracks}. This is a significant fraction of the total disc mass typically observed in post-AGB/post-RGB binaries \citep[$M_{\rm disc}\lesssim\,10^{-2}\,M_\odot$; see Table 4.7 in][]{gallardocava2023thesis}. Confirming this dissipation scenario will require direct analysis of the gas and dust distribution in faint discs using interferometric facilities, including Atacama Large Millimeter/submillimeter Array (ALMA) and Very Large Telescope Interferometer (VLTI).

In addition, the disc dissipation hypothesis is further supported by the case of BD+15\,2862 (see Section~\ref{ssec:sdosam}), a target with extremely low IR excess ($L_{\rm IR}/L_\ast < 0.01$) and a wide orbit ($P_{\rm orb}=530.9$\,d). Although excluded from our depletion analysis due to its high effective temperature ($T_{\rm eff}\sim12\,000$\,K; see Section~\ref{ssec:sdosam}), which precludes a homogeneous Fe line analysis with \texttt{E-iSpec}, BD+15\,2862 is known to be extremely depleted, with abundances determined for only seven elements, including [S/H] = --0.1\,dex and [Fe/H] $<$ --3\,dex \citep{giridhar2005HotFaint}. This system likely represents an advanced evolutionary stage along the post-AGB/post-RGB track, illustrating severe photospheric depletion in the final stages of CBD evolution.

We note that the faint disc target AU Vul (\#19) is an outlier in the disc dissipation scenario. This post-AGB binary is the coolest target in our combined sample \citep[$L\sim6550L_\odot$, $T_{\rm eff}\,=\,4740\,\pm\,40$\,K; see, e.g.,][]{kluska2022GalacticBinaries}. Although $T_{\rm eff}$ is not a perfect indicator of post-AGB age (due to potential stalling by accretion; see Section~\ref{ssec:dscall}), it is reasonable to infer that AU Vul (\#19) spent less time on the post-AGB track than other faint disc targets. Nevertheless, this target exhibits a significant depletion slope ($\nabla_{\rm 100\,K} = -0.17_{-0.10}^{+0.06}$\,dex), suggesting that disc dissipation and/or dust settling occurred unusually early or rapidly, possibly due to binary interaction, low initial disc mass, or an alternative dispersal mechanism. To better understand depletion in cool, chemically depleted post-AGB/post-RGB binaries, further homogeneous chemical analysis of similarly cool full disc targets are required. In addition, a homogeneous chemical analysis of all full disc targets in the Galaxy and the Magellanic Clouds will allow a robust test of our assumption that $T_{\rm turn-off}$ can serve as a discriminant between full and transition disc systems.

More generally, several physical mechanisms could accelerate disc dissipation and/or dust settling in post-AGB/post-RGB systems -- one possible contributor is planet formation. The growth of pebbles and planetesimals can generate additional pressure bumps that efficiently trap dust within the disc \citep{eriksson2020PebblesPlanetesimals, shibaike2023PebblesPlanetesimals, drazkowska2023DustPlanetFormation, price2025PebblesPlanetesimals}. The IR excesses in the SEDs of faint disc targets (see Figure~\ref{fig:allSED}) suggest that, if disc asymmetries or planet-induced spiral structures exist, they could be detectable with current interferometric facilities (e.g., ALMA and VLTI). Observing disc asymmetries in full, transition, and faint disc targets would provide critical insights into the role of planet formation in driving and modifying disc dissipation and dust-settling processes \citep{hillen2017DiscInterferometry, kluska2018IRAS08, kluska2019DiscSurvey, andrych2023Polarimetry, andrych2024IRAS08}.

To interpret this possibility, and to understand why faint and transition discs remain relatively rare among the post-AGB/post-RGB sample, current planet formation models for protoplanetary discs around young stars \citep[see, e.g.,][and references therein]{raymond2022PlanetFormationModels} need to be adapted to the unique conditions of CBDs around post-AGB/post-RGB binaries. This includes accounting for stellar irradiation, gas-to-dust ratios, and both chemical and spatial disc structures. Our ongoing observational and theoretical work will continue to investigate the evolution and mechanisms of photospheric depletion in these systems.

\section{Conclusions}\label{sec:con}
The aim of this study was to investigate the link between CBD evolution and photospheric depletion in post-AGB/post-RGB binaries with faint discs. For this aim, we performed a detailed chemical abundance analysis based on high-resolution optical spectra obtained with HERMES/Mercator and UVES/VLT. Using the \texttt{E-iSpec} code, we derived precise atmospheric parameters and elemental abundances for eight post-AGB/post-RGB binaries with faint discs in both the Galaxy and the LMC. In addition, we employed \texttt{pySME} to compute NLTE corrections for individual lines of a representative set of chemical elements from carbon to barium.

To consistently analyse the depletion profiles of faint disc targets, we fitted each profile with a two-piece linear function defined by three free parameters: initial metallicity [M/H]$_0$, turn-off temperature $T_{\rm turn-off}$, and depletion scale $\nabla_{\rm 100\,K}$. Our analysis of NLTE-corrected depletion profiles of faint disc targets revealed that these profiles are `saturated', with a weak bimodality in the $T_{\rm turn-off}$ distribution centered around $\sim$900\,K and $\sim$1\,300\,K, and a strong bimodality in the $\nabla_{\rm 100\,K}$ distribution clustered near $\sim-$0.45 and $\sim-$0.15\,dex (see blue distributions in the right panels of Figures~\ref{fig:dplteff}, \ref{fig:dpllum}, \ref{fig:dplporb}, and \ref{fig:dpllir}). Overall, [M/H]$_0$ spans from --1.17 to +0.25\,dex, $T_{\rm turn-off}$ ranges from 730 to 1\,280\,K, and $\nabla_{\rm 100\,K}$ extends from --0.53 to --0.11\,dex.

To place our results in context, we combined our faint disc sample with a previously studied subset of post-AGB/post-RGB binaries hosting full and transition discs. 
The distributions of [M/H]$_0$ and $\nabla_{\rm 100\,K}$ in the combined sample of post-AGB/post-RGB binaries appear stochastic, whereas the distribution of $T_{\rm turn-off}$ shows two distinct peaks at $\sim\,900$ and $\sim\,1\,300$\,K. Based on this bimodality, we categorised the faint disc sample into two subgroups: i) full-like targets and ii) transition-like targets. This categorisation suggests that different disc formation and evolution mechanisms may underlie the morphological differences in depletion profiles between the full and transition disc targets. The precise location of dust-gas separation within the disc, and its overlap with the disc region re-accreted onto the star, remain open questions. We note that our sample size is limited, and the observed trends may be affected by small-number statistics and selection biases. This work will be extended by our ongoing homogeneous abundance analysis of post-AGB/post-RGB binaries with full discs in the Galaxy and the Magellanic Clouds.

Our results confirm that post-AGB and post-RGB binaries with faint discs, despite their low IR excesses, retain strong photospheric depletion signatures, indicating that these systems previously hosted full or transition discs. Most faint disc targets (7 of 8) likely represent the final evolutionary stages of circumbinary disc dissipation, analogous to photoevaporation-driven clearing in protoplanetary systems. The cool faint disc target AU\,Vul (\#19) is an outlier, showing significant depletion despite being relatively less evolved, suggesting unusually rapid disc dissipation and/or dust settling. In our combined sample, we also find a significant anticorrelation between IR-to-stellar luminosity ratio and depletion slope in transition disc systems.

These findings highlight the need to investigate the gas and dust distribution in circumbinary discs by combining homogeneous abundance analyses of full, transition, and faint disc targets with high-resolution imaging and detailed physical–chemical modelling. The diversity of depletion profiles across these disc types underscores the complexity of disc–binary interactions in post-AGB/post-RGB binaries. To better constrain the processes driving photospheric depletion, we are expanding both our sample and methodological approach -- an essential step toward understanding the full diversity of depletion profiles in the post-AGB/post-RGB binary population.

\section*{Acknowledgements}\label{sec:ack}
The spectroscopic results presented in this study are based on observations obtained with the HERMES spectrograph, which is supported by the Research Foundation - Flanders (FWO), Belgium, the Research Council of KU Leuven, Belgium, the Fonds National de la Recherche Scientifique (F.R.S.-FNRS), Belgium, the Royal Observatory of Belgium, the Observatoire de Gen\`{e}ve, Switzerland and the Th\"{u}ringer Landessternwarte Tautenburg, Germany. This research is also based on observations collected at the European Organisation for Astronomical Research in the Southern Hemisphere under ESO programmes 074.D-0619 and 092.D-0485. This research was supported by computational resources provided by the Australian Government through the National Computational Infrastructure (NCI) under the National Computational Merit Allocation Scheme and the ANU Merit Allocation Scheme (project y89).

MM1 acknowledges the International Macquarie Research Excellence Scholarship program (iMQRES) for financial support during the research. MM1, DK, HVW, and KA acknowledge the support of the Australian Research Council Discovery Project DP240101150. AMA acknowledges support from the Swedish Research Council (VR 2020-03940), the Crafoord Foundation via the Royal Swedish Academy of Sciences (CR 2024-0015), and the European Union’s Horizon Europe research and innovation programme under grant agreement No. 101079231 (EXOHOST).

\section*{Data Availability Statement}
The data underlying this article are available in the article and in its online supplementary material.

% \defbibnote{preamble}{By default, this template uses \texttt{biblatex} and adopts the Chicago referencing style. However, the journal you’re submitting to may require a different reference style; specify the journal you're using with the class' \texttt{journal} option --- see lines 1--8 of \emph{sample.tex} for a list of options and instructions for selecting the journal.}

\bibliography{master}%[prenote={preamble}]
\appendix

\section{Additional notes on selected targets}\label{app:tar}
In this Appendix, we provide supplementary information on the circumstellar gas and dust characteristics of several targets, based on previous studies not included in Table~\ref{tab:litpar}. Specifically, we include additional notes for SS Gem, V382 Aur, R Sct, and BD+39\,4926.

\textbf{SS Gem.} The RV phase curve of this target displays a long-term modulation of the mean brightness, superimposed on regular pulsations, known as RVb phenomenon (see Table~\ref{tab:litpar}). This additional variability is attributed to binarity and variable obscuration by circumstellar dust along the line of sight \citep{kiss2017RVTau}.

\textbf{V382 Aur.} This target displays an RVb phenomenon in the RV phase curve (see Table~\ref{tab:litpar}) and emission from carbonaceous dust in the IR spectra from Spitzer \citep[polycyclic aromatic hydrocarbons and C$_{60}$ fullerenes;][]{gielen2011CarbonaceousDust}.

\textbf{R Sct.} The distribution of gas and dust around this target was mapped in $^{12}$CO J=2-1 using IRAM NOrthern Extended Millimeter Array \citep[NOEMA;][]{gallardocava2021PostAGBOutflows}. With a NOEMA beam size of 3.12''\,$\times$\,2.19'', CBD in R\,Sct was unresolved ($R\,\lesssim\,700$\,au), though the large outflow was detected ($\sim\,8\,000\,$x$\,15\,000$\,au).

\textbf{BD+39\,4926.} This target was proposed to host a dissipated disc based on low IR excess in the SED and high effective temperature \citep{gezer2015WISERVTau}.

\section{Summary of optical spectral visits}\label{app:vis}
In this Appendix, we list all optical spectral visits considered in this study (see Table~\ref{tabA:obslog}) for the abundance analysis of post-AGB/post-RGB binaries with faint discs (see Section~\ref{ssec:anaspc}).

\begin{table}[!ht]
    \centering
    \scriptsize
    \caption[Spectral visits of post-AGB/post-RGB binaries with faint discs]{Spectral visits of post-AGB/post-RGB binaries with faint discs (see Section~\ref{ssec:sdospc}). This table is published in its entirety in the electronic edition of the paper. A portion is shown here for guidance regarding its form and content.}\label{tabA:obslog}
    \begin{tabular}{|c|c|c|c|}
    \hline
        \textbf{ObsID} & \textbf{MJD} & \textbf{Phase} & \textbf{RVrel} \\ \hline
        \multicolumn{4}{|c|}{\textit{SS Gem (RV0=-8.4 km/s, 151 visits)}} \\ \hline
        253252 & 55124.23258 & 0.20 & 2.8 \\
        260288 & 55160.06403 & 0.00 & -13.6 \\
        272381 & 55214.99618 & 0.23 & -3.0 \\
        273153 & 55221.92900 & 0.39 & -1.1 \\
        273787 & 55232.98736 & 0.63 & -7.3 \\
        \multicolumn{4}{|c|}{\ldots} \\ \hline
    \end{tabular}\\
    \textbf{Notes:} RV0 is the radial velocity of the spectral visit with the best S/N ratio. RVrel are radial velocities relative to visit with the best S/N ratio.
    %\vspace{-0.75cm}
\end{table}

\section{Master line list of the target sample}\label{app:lst}
In this Appendix, we show the combined optical line list (see Table~\ref{tabA:linlst}), which we used to derive the atmospheric parameters and elemental abundances of post-AGB/post-RGB binaries with faint discs (see Section~\ref{ssec:anaspc}).

\begin{table}[!ht]
    \centering
    \scriptsize
    \caption[Line list of the whole studied sample of post-AGB/post-RGB binaries with full, transition, and faint discs (see Section~\ref{sec:dsc})]{Line list of the whole studied sample of post-AGB/post-RGB binaries with full, transition, and faint discs (see Section~\ref{sec:dsc}). This table is published in its entirety in the electronic edition of the paper. A portion is shown here for guidance regarding its form and content.}\label{tabA:linlst}
    \resizebox{\textwidth}{!}{
    \begin{tabular}{|c@{\hspace{0.1cm}}ccc|cccc|} \hline
        \multicolumn{4}{|c|}{\textbf{Atomic data}} & \multicolumn{4}{c|}{\boldmath$W_\lambda$ \textbf{(m\AA)}} \\
        \textbf{Ion} & \boldmath$\lambda$ & \boldmath$\log gf$ & \boldmath$\chi$ & \textbf{SZ Mon} & \textbf{DF Cyg} & \textbf{\ldots} & \textbf{J053254} \\
        ~ & \textbf{(nm)} & \textbf{(dex)} & \textbf{(eV)} & \textbf{(\#1)} & \textbf{(\#2)} & \textbf{\ldots} & \textbf{(\#22)} \\ \hline
        \ion{C}{i} & 402.9414 & -2.135 & 7.488 & -- & -- & \ldots & -- \\ 
        \ion{C}{i} & 422.8326 & -1.914 & 7.685 & -- & -- & \ldots & -- \\
        \ion{C}{i} & 426.9019 & -1.637 & 7.685 & -- & -- & \ldots & -- \\
        \ion{C}{i} & 437.1367 & -1.962 & 7.685 & -- & -- & \ldots & -- \\
        \ion{C}{i} & 477.0021 & -2.437 & 7.483 & -- & -- & \ldots & -- \\
        \multicolumn{8}{|c|}{\ldots} \\ \hline
    \end{tabular}
    }
    %\vspace{-0.75cm}
\end{table}

\section{Individual abundances of faint disc targets}\label{app:abu}
In this Appendix, we present the derived LTE and NLTE elemental abundances of faint disc targets (see Tables~\ref{tabA:reslte} and \ref{tabA:resnlt}, respectively). For more details on the abundance analysis, see Section~\ref{ssec:anaspc}.

\begin{table}[!ht]
    \centering
    \scriptsize
    \caption[LTE {[X/H]} abundances of the whole studied sample of post-AGB/post-RGB binaries with full, transition, and faint discs (see Section~\ref{sec:dsc})]{LTE [X/H] abundances of the whole studied sample of post-AGB/post-RGB binaries with full, transition, and faint discs (see Section~\ref{sec:dsc}).}\label{tabA:reslte} % This table is published in its entirety in the electronic edition of the paper. A portion is shown here for guidance regarding its form and content.
    \begin{tabular}{|c|c|cccc|}
    \hline
        \textbf{Ion} & \boldmath$T_{\rm cond}$ & \textbf{SZ Mon} & \textbf{DF Cyg} & \textbf{\ldots} & \textbf{J053254} \\
        ~ & \textbf{(K)} & \textbf{(\#1)} & \textbf{(\#2)} & \textbf{\ldots} & \textbf{(\#22)} \\ \hline
        %\multicolumn{7}{|c|}{\textit{LTE}} \\ \hline
        \ion{C}{i} & 40 & --0.09$\pm$0.08 & 0.27$\pm$0.09 & \ldots & --0.42$\pm$0.06 \\
        \ion{N}{i} & 123 & 0.44$\pm$0.14 & -- & \ldots & -- \\
        \ion{O}{i} & 183 & 0.15$\pm$0.11 & 0.73$\pm$0.12 & \ldots & --0.63$\pm$0.12 \\
        \ldots & \ldots & \ldots & \ldots & \ldots & \ldots \\ \hline
    \end{tabular}\\
    \textbf{Note:} The condensation temperatures for C and N were adopted from \cite{lodders2003CondensationTemperatures}, the condensation temperatures for all other elements were adopted from \cite{wood2019CondensationTemperatures}.
    %\vspace{-0.75cm}
\end{table}

\begin{table}[!ht]
    \centering
    \scriptsize
    \caption[NLTE {[X/H]} abundances of the whole studied sample of post-AGB/post-RGB binaries with full, transition, and faint discs (see Section~\ref{sec:dsc})]{NLTE [X/H] abundances of the whole studied sample of post-AGB/post-RGB binaries with full, transition, and faint discs (see Section~\ref{sec:dsc}).}\label{tabA:resnlt} % This table is published in its entirety in the electronic edition of the paper. A portion is shown here for guidance regarding its form and content.
    \begin{tabular}{|c|c|cccc|}
    \hline
        \textbf{Ion} & \boldmath$T_{\rm cond}$ & \textbf{SZ Mon} & \textbf{DF Cyg} & \textbf{\ldots} & \textbf{J053254} \\
        ~ & \textbf{(K)} & \textbf{(\#1)} & \textbf{(\#2)} & \textbf{\ldots} & \textbf{(\#22)} \\ \hline
        %\multicolumn{7}{|c|}{\textit{NLTE}} \\ \hline
        \ion{C}{i} & 40 & --0.15$\pm$0.08 & 0.14$\pm$0.07 & \ldots & --0.48$\pm$0.06 \\
        \ion{N}{i} & 123 & 0.20$\pm$0.14 & -- & \ldots & -- \\
        \ion{O}{i} & 183 & 0.15$\pm$0.11 & 0.84$\pm$0.05 & \ldots & --0.62$\pm$0.12 \\
        \ldots & \ldots & \ldots & \ldots & \ldots & \ldots \\ \hline
    \end{tabular}\\
    \textbf{Note:} The condensation temperatures for C and N were adopted from \cite{lodders2003CondensationTemperatures}, the condensation temperatures for all other elements were adopted from \cite{wood2019CondensationTemperatures}.
    %\vspace{-0.75cm}
\end{table}

\section{NLTE depletion profiles of full and transition disc targets}\label{app:dpl}
\begin{figure}[!hb]
    \centering
    \includegraphics[width=.81\linewidth]{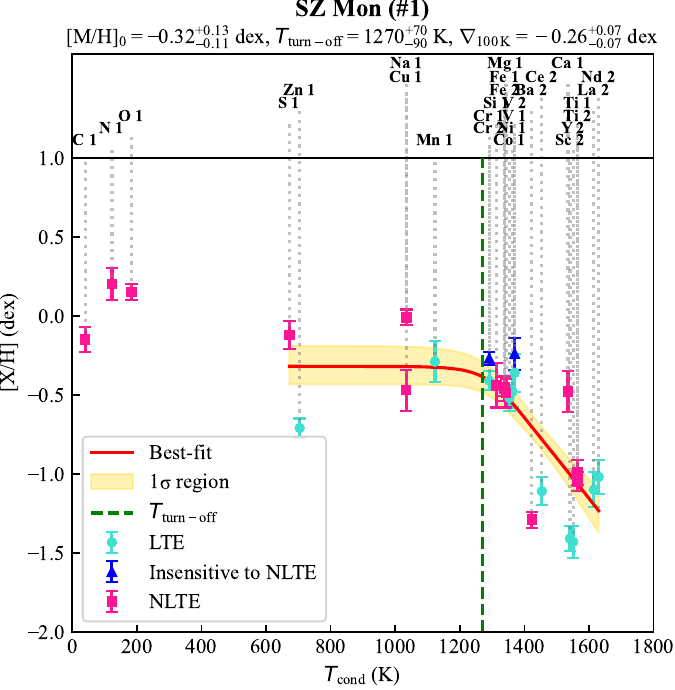}
    \includegraphics[width=.81\linewidth]{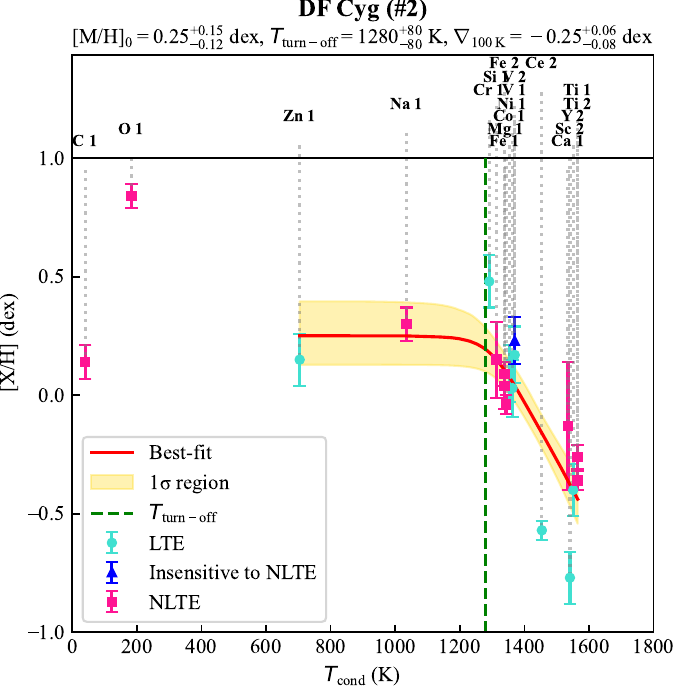}
    \caption[Elemental abundances of a subsample of post-AGB/post-RGB binaries with full discs as functions of condensation temperature]{Elemental abundances of a subsample of post-AGB/post-RGB binaries with full discs as functions of condensation temperature \citep{lodders2003CondensationTemperatures, wood2019CondensationTemperatures}. The legend for the symbols and colours used is included within the plot. ``NLTE insensitive'' abundances are derived from spectral lines of \ion{V}{ii} and \ion{Cr}{ii}; for more details, see Section~\ref{ssec:anaspc}).}\label{figA:dpl1}
    \vspace{-0.25cm}
\end{figure}

\begin{figure*}
    \centering
    \includegraphics[width=.405\linewidth]{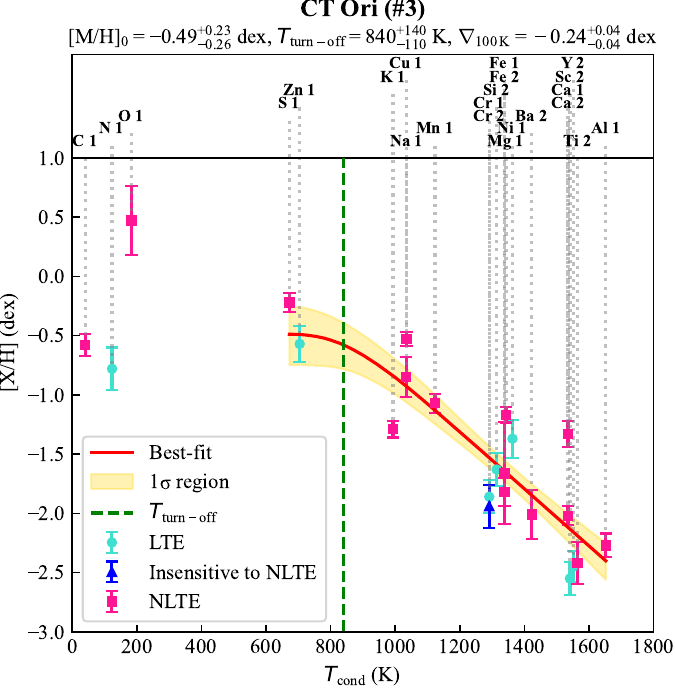}
    \includegraphics[width=.405\linewidth]{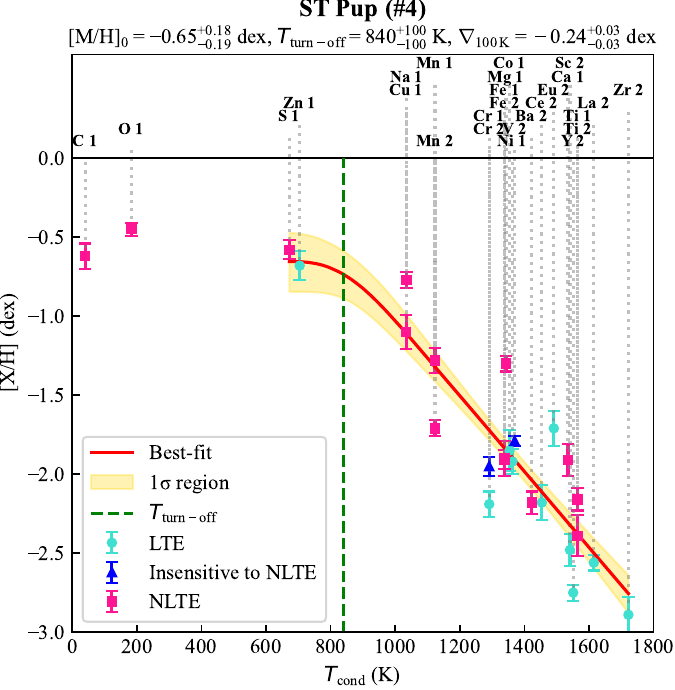}
    \includegraphics[width=.405\linewidth]{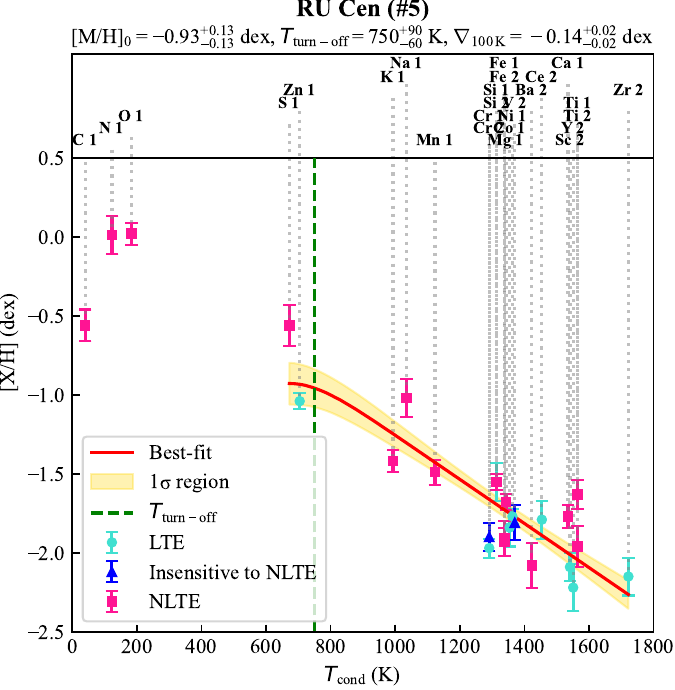}
    \includegraphics[width=.405\linewidth]{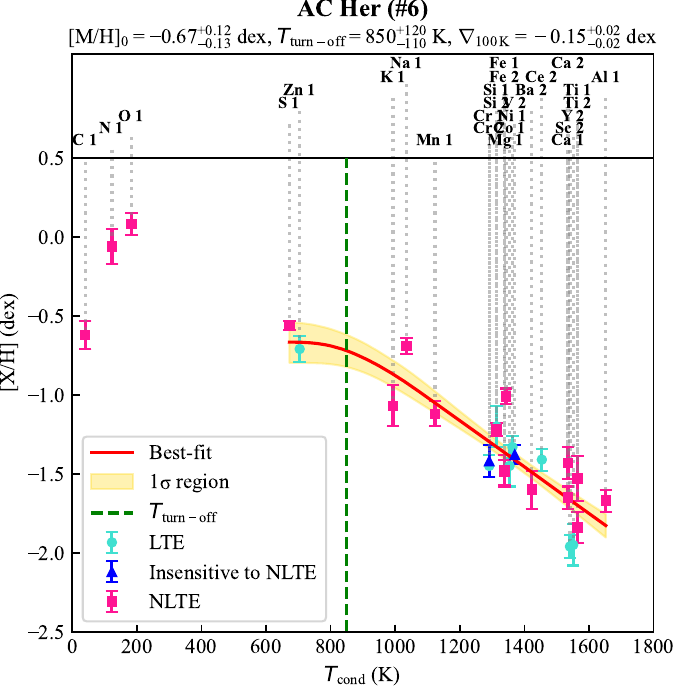}
    \includegraphics[width=.405\linewidth]{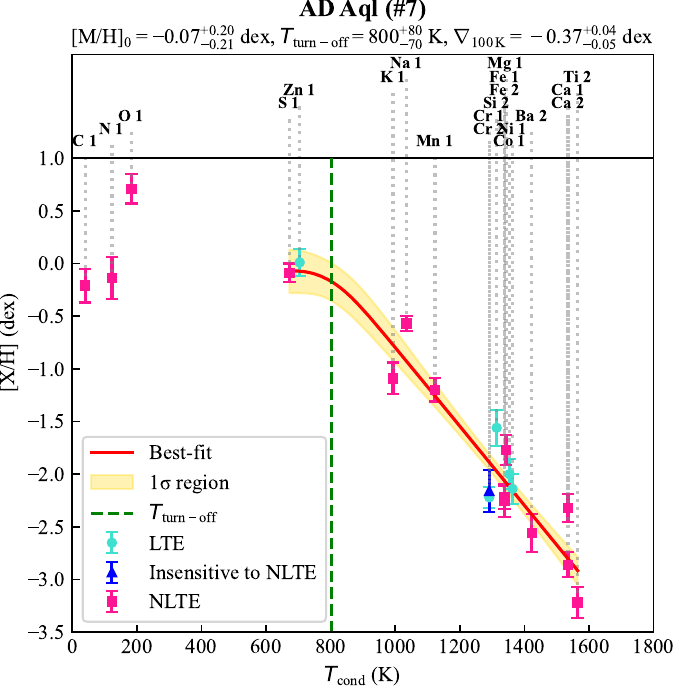}
    \includegraphics[width=.405\linewidth]{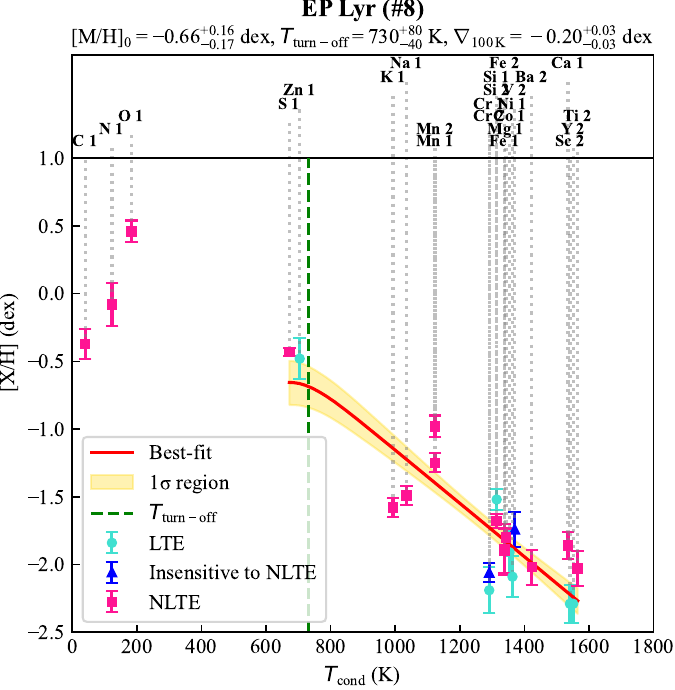}
    \caption[Elemental abundances of a subsample of post-AGB/post-RGB binaries with full and transition discs as functions of condensation temperature]{Elemental abundances of a subsample of post-AGB/post-RGB binaries with full and transition discs as functions of condensation temperature \citep{lodders2003CondensationTemperatures, wood2019CondensationTemperatures}. The legend for the symbols and colours used is included within the plot. ``NLTE insensitive'' abundances are derived from spectral lines of \ion{V}{ii} and \ion{Cr}{ii}; for more details, see Section~\ref{ssec:anaspc}).}\label{figA:dpl2}
\end{figure*}

\begin{figure*}
    \centering
    \includegraphics[width=.405\linewidth]{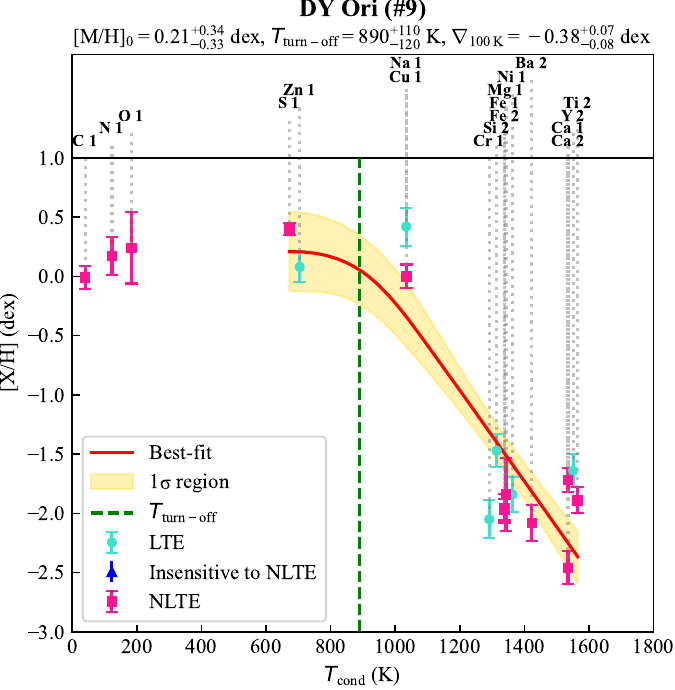}
    \includegraphics[width=.405\linewidth]{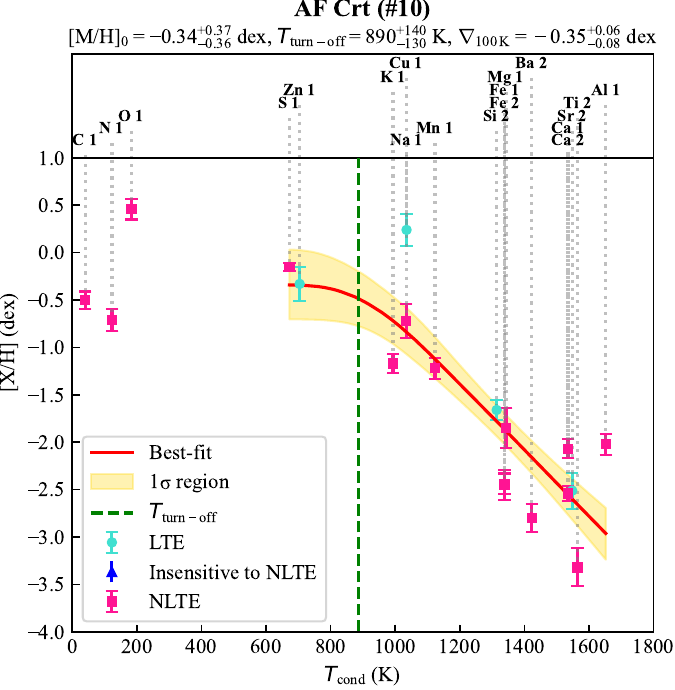}
    \includegraphics[width=.405\linewidth]{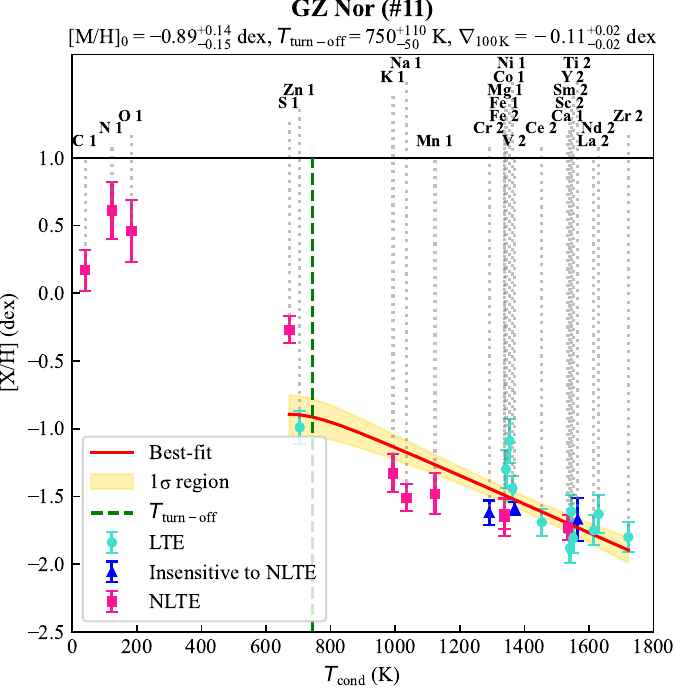}
    \includegraphics[width=.405\linewidth]{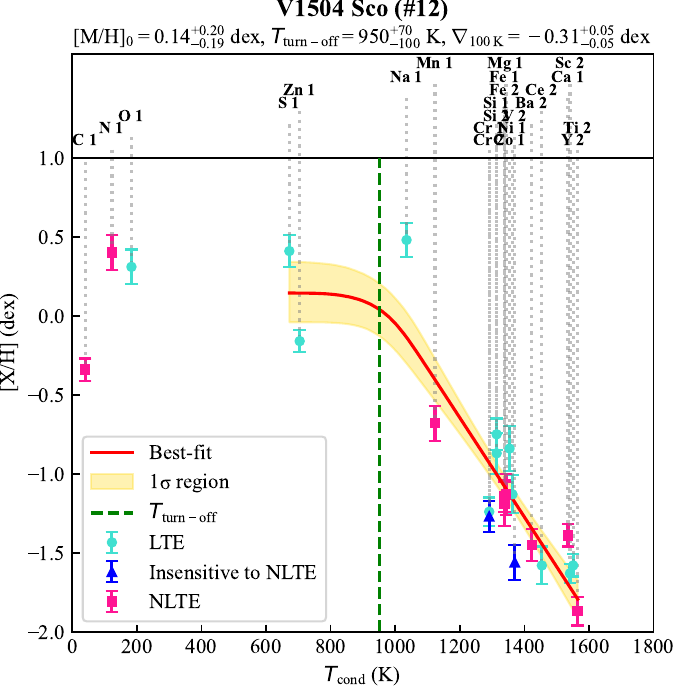}
    \includegraphics[width=.405\linewidth]{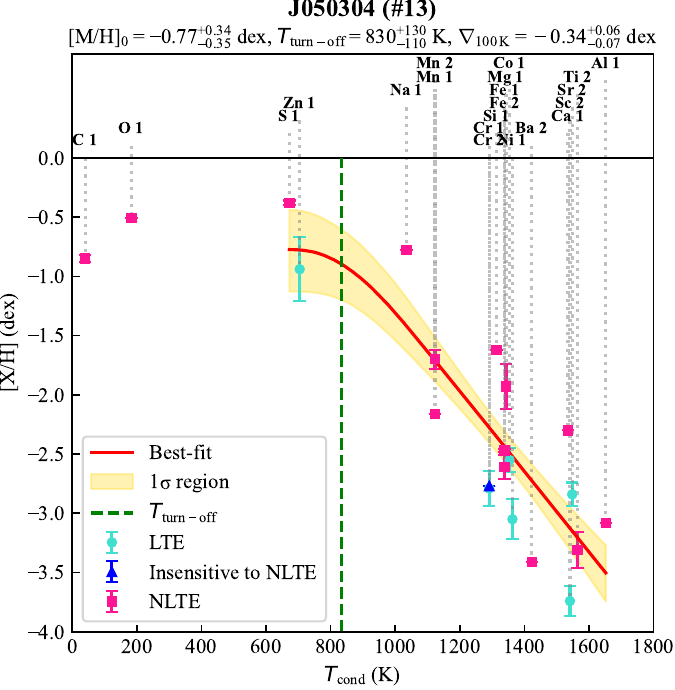}
    \includegraphics[width=.405\linewidth]{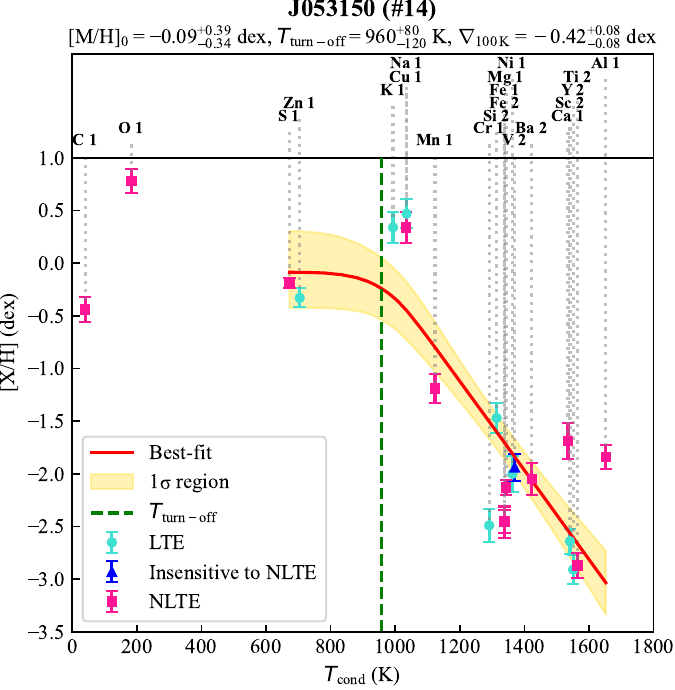}
    \caption[Elemental abundances of a subsample of post-AGB/post-RGB binaries with transition discs as functions of condensation temperature]{Elemental abundances of a subsample of post-AGB/post-RGB binaries with transition discs as functions of condensation temperature \citep{lodders2003CondensationTemperatures, wood2019CondensationTemperatures}. The legend for the symbols and colours used is included within the plot. ``NLTE insensitive'' abundances are derived from spectral lines of \ion{V}{ii} and \ion{Cr}{ii}; for more details, see Section~\ref{ssec:anaspc}).}\label{figA:dpl3}
\end{figure*}

In this Appendix, we show the revised depletion profiles of full and transition disc targets (see Figures~\ref{figA:dpl1}, \ref{figA:dpl2}, \ref{figA:dpl3}). To ensure consistency across the full, transition, and faint disc samples, we adopted the LTE abundances and luminosities of the full and transition disc targets from \citet{mohorian2024EiSpec} and \citet{mohorian2025TransitionDiscs}, which were derived using the same methodology as in this study (see Section~\ref{ssec:anaspc}). However, for a uniform NLTE analysis across all targets, we recalculated NLTE corrections for full and transition disc samples using \texttt{pySME}. These new corrections were applied to the published LTE abundances using the same procedure as for the faint disc sample (see Section~\ref{ssec:ananlt}). For more details on the depletion profiles of full and transition disc targets, see Section~\ref{ssec:dscall}.

We note that for transition discs, the NLTE abundances derived in this study generally agree with those presented in \citet{mohorian2025TransitionDiscs} within typical uncertainties of $\sim$0.15 dex, with  a median difference of 0.07\,dex and an average difference of 0.10\,dex. These discrepancies largely arise from two factors: i) the need to interpolate or extrapolate NLTE corrections near the edges of the MARCS grid, which is irregular and incomplete in the parameter space relevant to post-AGB/post-RGB binaries, and ii) adoption of different approaches for estimating NLTE corrections \citep[][matched LTE and NLTE equivalent widths; in this study, we matched LTE and NLTE synthetic spectra]{mohorian2025TransitionDiscs}. We also note that within the post-AGB/post-RGB sample, the average NLTE corrections for volatile elements (including C, N, O, and S) are $\approx\,-0.07$\,dex, whereas the average NLTE corrections for refractory elements (including Al, Si, Ti, and Fe) are $\approx\,+0.08$\,dex. This confirms that, while NLTE corrections are essential for deriving precise individual abundances, these corrections have a minor impact on the general distributions of depletion parameters.

\end{document}